\documentclass[useAMS,usenatbib]{mn2e}
\usepackage{graphicx}

\title[The X-ray Properties of Mrk~59 and Mrk~71]
{The X-ray Properties of the Cometary Blue Compact Dwarf galaxies Mrk~59 
and Mrk~71}
\author[T. X. Thuan et al.]{T.~X.~Thuan$^{1}$,
F.~E.~Bauer$^{2,3,4}$ and Y.~I.~Izotov$^{5}$
\\
$^{1}$Astronomy Department, University of Virginia,
P.O. Box 3818, University Station, Charlottesville, VA, 22903, USA;
txt@virginia.edu \\ 
$^{2}$Instituto de Astrof\'{\i}sica, Facultad de F\'{i}sica, Pontificia 
Universidad Cat\'{o}lica de Chile, Casilla 306, Santiago 22, Chile; 
fbauer@astro.puc.cl \\
$^{3}$Millennium Institute of Astrophysics\\
$^{4}$Space Science Institute, 4750 Walnut Street, Suite 205, 
Boulder, CO 80301, USA; fbauer@spacescience.org \\
$^{5}$Main
Astronomical Observatory, National Academy of Sciences of Ukraine,
03680 Kyiv, Ukraine; izotov@mao.kiev.ua}
\begin{document}

\date{07 April 2014}

\pagerange{\pageref{firstpage}--\pageref{lastpage}} \pubyear{2012}

\maketitle

\label{firstpage}

\begin{abstract}
We present {\sl XMM-Newton} and {\sl Chandra} observations of two
low-metallicity cometary blue compact dwarf (BCD) galaxies, Mrk 59 and
Mrk~71.  The first BCD, Mrk 59, contains two ultraluminous X-ray (ULX)
sources, IXO 72 and IXO 73, both associated with bright massive stars
and H {\sc ii} complexes, as well as one fainter extended source
associated with a massive H {\sc ii} complex at the head of the
cometary structure.  The low-metallicity of Mrk 59 appears to be
responsible for the presence of the two ULXs.  IXO 72 has varied little
over the last 10 yr, while IXO 73 has demonstrated a
variability factor of $\sim$4 over the same period.  The second BCD,
Mrk 71, contains two faint X-ray point sources and two faint extended
sources.  One point source is likely a background AGN, while the other
appears to be coincident with a very luminous star and a compact H
{\sc ii} region at the ``head'' of the cometary structure.  The two
faint extended sources are also associated with massive H {\sc ii}
complexes.  Although both BCDs have the same metallicity, the three
sources in Mrk 71 have X-ray luminosities $\sim$1--2 orders of
magnitude fainter than those in Mrk 59. The age of the starburst may
play a role.
\end{abstract}


\section{Introduction}\label{intro}

With a heavy element abundance ranging from 3\% to 50\% that of the
Sun, blue compact dwarf (BCD) galaxies are the least chemically
evolved gas-rich star-forming galaxies known in the local universe
\citep{T08}.  They thus constitute the best laboratories for studying
physical processes which occurred at high redshifts, when the gas was
very metal-deficient.

BCDs are undergoing intense bursts of star formation, giving birth to
thousands of O stars in a very compact starburst region.  Because of
the presence of these many massive short-lived stars, BCDs are
expected to emit in the X-ray.
This X-ray emission can come from compact sources such as high-mass
X-ray binaries (HMXBs) and/or hot O and Wolf-Rayet stars, or from
diffuse sources such as the hot plasma associated with supernova (SN)
remnants. Stellar winds and SNe inject energy and momentum into the
cold ambient interstellar medium (ISM), producing large amounts of hot
gas.  The starburst activity that injects hot X-ray-emitting gas into
the ISM lasts about 10$^7$ yr.  This starburst phase is then followed
by a long ($>$ 1--2 Gyr) quiescent period of passive photometric
evolution, before the occurrence of the next burst.  Depending on the
energy injection rate into the ISM and the geometry and robustness of
the cold gaseous ambient medium, expansion of the hot ISM on scales
comparable to the galactic scale length can result in a funneling of
hot gaseous mass into the cold gaseous halo. Because of the relatively
low potential well of BCDs, mass loss can occur \citep{DH94}.
Extensive mass loss can then lead to an expansion of the galaxy's size
and to a morphological evolution of the dwarf galaxy \citep{YA87}.

We present in this paper an X-ray study of two BCDs, Mrk 59$\equiv$I
Zw 49 and Mrk 71$\equiv$NGC 2363.  These are of particular interest
because they are the
prototypes of a particular class of BCDs, dubbed ``cometary'' galaxies
by \citet{LT85} in their BCD morphological classification
scheme. Cometary BCDs are characterized by a high surface brightness
star-forming region (the comet's ``head'') at one end of an elongated
low surface brightness stellar body (the comet's ``tail''), suggestive
of a flattened dwarf irregular galaxy seen nearly edge-on.  In the
case of Mrk 59, the irregular galaxy is called NGC 4861, and in the
case of Mrk 71, NGC 2366.  In the following, we will be using the
names of Mrk 59 and Mrk 71 to designate {\it both} the high surface
brightness star-forming region and the low surface brightness
elongated stellar body.  The bright star-forming region is at the end
of a long chain of fainter and smaller H~{\sc ii} regions embedded in
the lower surface brightness stellar body of the cometary BCD.  This
chain of H~{\sc ii} regions is suggestive of self-propagating star
formation which stopped at the edge of the galaxy.  By studying
cometary BCDs, we can examine how star formation ignites and
propagates in low-mass gas-rich stellar systems. Because the brightest
H~{\sc ii} region at the end of the chain is youngest and those along
the chain are progressively older with increasing distance from the
edge \citep{N00}, we can study the time evolution of H~{\sc ii} regions, and in
particular the time dependence of their X-ray properties.

Of the 2 BCDs, only Mrk 59 has been observed before in the X-ray
range.  Using {\sl Einstein}, \citet{F92} detected it as a strong
X-ray source with an X-ray luminosity of $\sim$10$^{40}$ erg
s$^{-1}$. Subsequently, \citet{P98} found from {\sl ROSAT} HRI
observations of Mrk 59 that the X-ray emission splits into two sources
separated by $\sim$35\arcsec. The southern one appears to coincide
with the high surface brightness starburst region at the end of the
stellar body, while the more luminous (by a factor of $\sim$2.5)
northern one is in the low surface brightness main body, and did not
appear to be associated with any evident H~{\sc ii} region.
  
The two BCDs have also been studied extensively at other wavelength
ranges.  Abundance determinations give oxygen abundances 12 + log O/H
of 8.0 and 7.9 for Mrk 59 and Mrk 71, respectively \citep{N00},
corresponding to 1/5 and 1/6 of the Sun's metallicity, if the solar
calibration (12 + log O/H)$_\odot$ = 8.7 of \citet{A09} is
adopted. \citet{N00} have derived O abundances for two other emission
knots along the elongated body of Mrk 59 and found them to be the same
as that of the bright knot, within the errors. The small scatter in
metallicity along the major axis of Mrk 59 ($\sim$0.2 dex) suggests
that the mixing of elements in the ionized gas has been efficient on a
spatial scale of several kiloparsecs. As for Mrk 71, \citet{R96} found
that the O abundance in several other H~{\sc ii} regions in the main
body varies between 8.1 and 8.3, slightly higher than in the brightest
H~{\sc ii} region.

\citet{T02} have used the {\sl Far~Ultraviolet~Spectroscopic~Explorer}
({\sl FUSE}) to study the abundances in the neutral ISM of Mrk 59 from
UV absorption lines.  They found that the heavy element abundance in
the neutral gas of Mrk 59 is about a factor of 10 less than that of
the ionized gas, or about 1/50 of the solar abundance. Although it has
a very low metallicity, the neutral gas of Mrk 59 is not pristine and
must have been enriched by previous generations of stars.
Using photometric and spectroscopic observations, \citet{N00} found
that the age of the oldest stars in the low surface brightness
component probably does not exceed $\sim$ 4 Gyr in Mrk 59 and $\sim$ 3
Gyr in Mrk71.  This age is smaller than the typical age (5 Gyr or
greater) of the underlying stellar population in BCDs of other types.
Cometary galaxies
thus appear to be relatively young galaxies. 

\citet {TI05} have used {\sl HST} $V$ and $I$ images to perform a
color-magnitude diagram (CMD) analysis of the stellar populations in
Mrk 71. The CMD reveals not only young stellar populations such as
blue main sequence stars (age $\la$ 30 Myr), but also an
intermediate-age population of blue and red supergiants (20 Myr $\la$
age $\la$ 100 Myr), and an older evolved stellar population of
asymptotic giant branch (AGB) stars (age $\ga$100 Myr) and red giant
stars (age $\ga$ 1 Gyr). This suggests that, in addition to the
present burst with age $\la$ 100 Myr, star formation in Mrk 71 has
started some 3 Gyr ago, consistent with the photometric age estimate
of \citet{N00}.  Near-infrared molecular hydrogen emission has been
detected in both Mrk 59 \citep{I09} and Mrk 71
\citep{I11}. \citet{THL04} have studied the H {\sc i} distribution and
kinematics of the two BCDs. The VLA maps show multiple H {\sc i} peaks
scattered over the disk. The latter shows regular rotational
kinematics, with a linear rise followed by a flattening of the
rotation curve.
     
In this paper, we will adopt a distance of 10.7~Mpc for Mrk 59
\citep{THL04}.
%
As for Mrk 71, we will use the Cepheid-derived distance of 3.44~Mpc 
\citep{T95}, placing it in the M81 group.  
At those distances, 1\arcsec\ corresponds to a linear size of 52 pc in
Mrk 59 and of 17 pc in Mrk 71. The Galactic column densities for
Mrk~59 and Mrk~71 are $N_{\rm H}=1.2\times10^{20}$~cm$^{-2}$ and
$N_{\rm H}=4.0\times10^{20}$~cm$^{-2}$, respectively, although based
on the H {\sc i} maps of \citet{THL04}, the internal neutral hydrogen
column densities in Mrk~59 and Mrk~71 could be as large as $N_{\rm
  H}=3\times10^{21}$~cm$^{-2}$.

\section{Observations and Data Reduction}\label{reduction}
 
\subsection{{\sl XMM--Newton} Observations}\label{xray}

Mrk~59 and Mrk~71 were observed by the {\sl XMM-Newton
    Observatory} with the $pn$--CCD camera \citep{Struder2001} and the
  two MOS--CCD cameras \citep{Turner2001} using the medium filter in
  full field mode. Mrk 59 was observed on three separate occasions
  during 2003 June 14, 2003 July 10, and 2003 December 3 (PI: Thuan;
  ObsIds 0141150101, 0141150401, and 0141150501, respectively;
  hereafter epochs 1 (E1), 2 (E2), and 3 (E3)). Mrk~71 was observed on 2002
  October 31 (PI: Thuan, ObsId 0141150201). The galaxies were
  generally placed near the aimpoint, allowing the entire optical
  extent of the galaxy to be imaged easily. The processing, screening,
  and analysis of the data were performed using the standard tools
  from {\tt SAS} (v.13.0.0), as well as custom {\tt IDL} software. The
  raw $pn$ and MOS data were initially processed using the standard
  {\it epchain} and {\it emchain} pipeline scripts. Time intervals
  contaminated by soft-proton flares were identified using the
  background light curve in the 10--15~keV band. After excluding
  background flares, our final exposures with the $pn$ (MOS) detectors 
were 13.1~ks (14.5~ks),
  4.0~ks (8.1~ks), 0.0~ks (9.8~ks) for epochs E1, E2, and E3 of Mrk 59, 
respectively, and
  29~ks (39~ks) for Mrk~71. We selected only good event patterns for further
  study: $\le$12 for MOS imaging and spectroscopy, $\le$4 for $pn$
  imaging and spectroscopy.

X-ray source detection was performed on the MOS and $pn$ images in the
0.3--10~keV band using the standard {\tt SAS} {\it eboxdetect} and
{\it emldetect} algorithms. Four X-ray sources were found to be
coincident with each of the optical extent of Mrk 59 and Mrk 71.
Background-subtracted counts were extracted using circular apertures
with radii in the range \hbox{$\approx$15--30\arcsec}, depending on
whether the source appeared point-like or extended. Local
backgrounds were determined from annuli after removal of nearby point
sources. Our background-subtracted detection limit corresponds to
$\approx 15$ $pn$ counts, or absorbed 0.5--10~keV fluxes of $\approx
2.8\times10^{-15}$~erg~cm$^{-2}$~s$^{-1}$ and $\approx
1.4\times10^{-15}$~erg~cm$^{-2}$~s$^{-1}$ in Mrk~59 and Mrk~71,
respectively, assuming a photon index $\Gamma$$=2$ and Galactic column densities.

For the three brightest {\sl XMM-Newton} sources,
XMMU~J125901.7+345115 and XMMU~J125900.7+345048 in Mrk~59 and
XMMU~J072857.9+691135 in Mrk~71, we extracted MOS and $pn$
  point-source spectra. We chose relatively small apertures
  (15--20\arcsec, corresponding to encircled energy fractions of
  $\approx0.67$--$0.74$); these included most of the counts, but
  were still small enough to avoid potential contamination in the case
  of Mrk~59, and limit background contamination for faint sources.
Local backgrounds were chosen to be on the same
chip. Event PI values and photon energies were determined using the
latest gain files appropriate for the observation, and spectral
products were generated using standard methods within {\tt SAS}.

\begin{figure*}
\vspace{-0.1in}
\begin{center}
\includegraphics[height=5.3cm, angle=0]{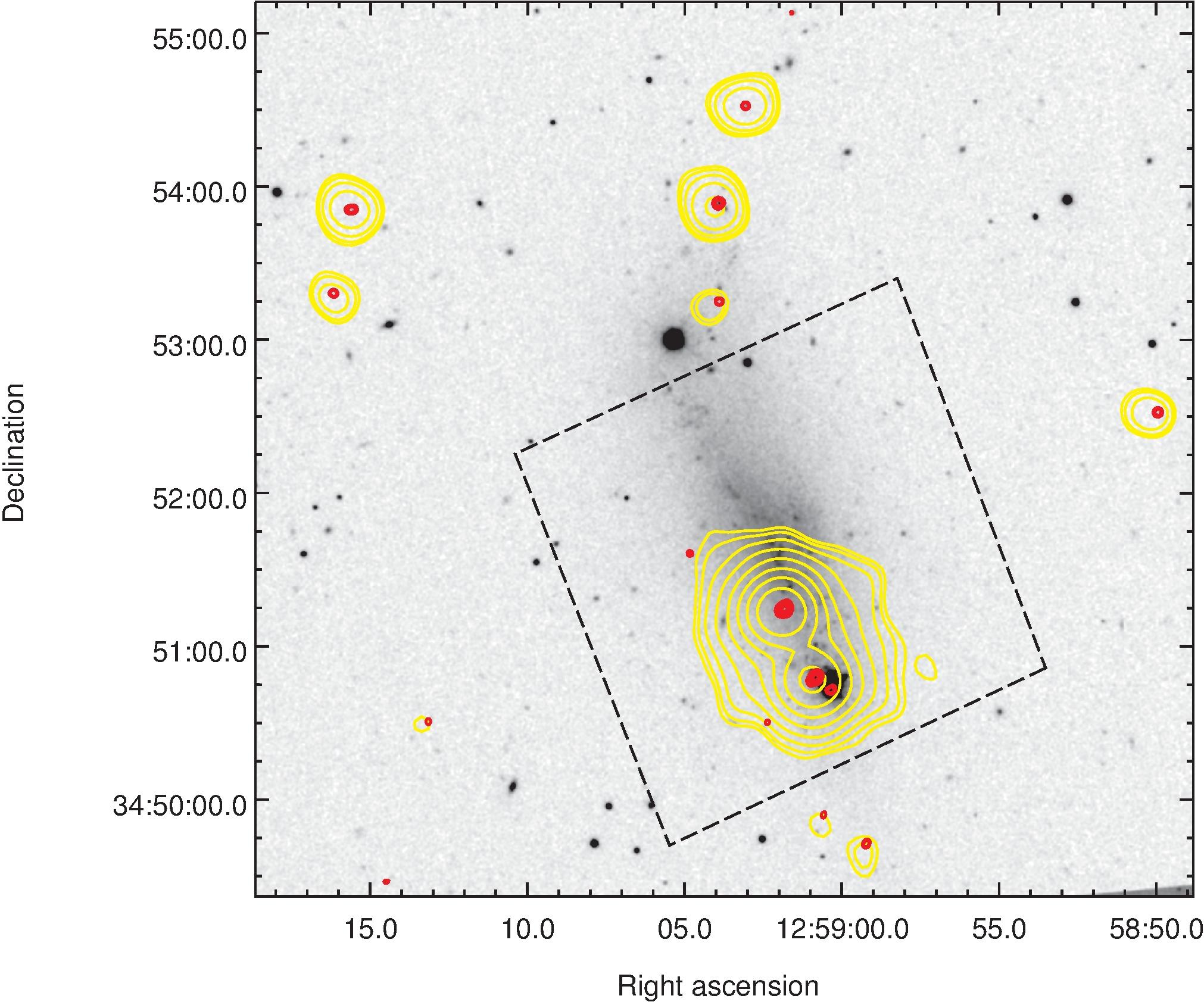}
\hglue-0.0in{\includegraphics[height=5.3cm, angle=0]{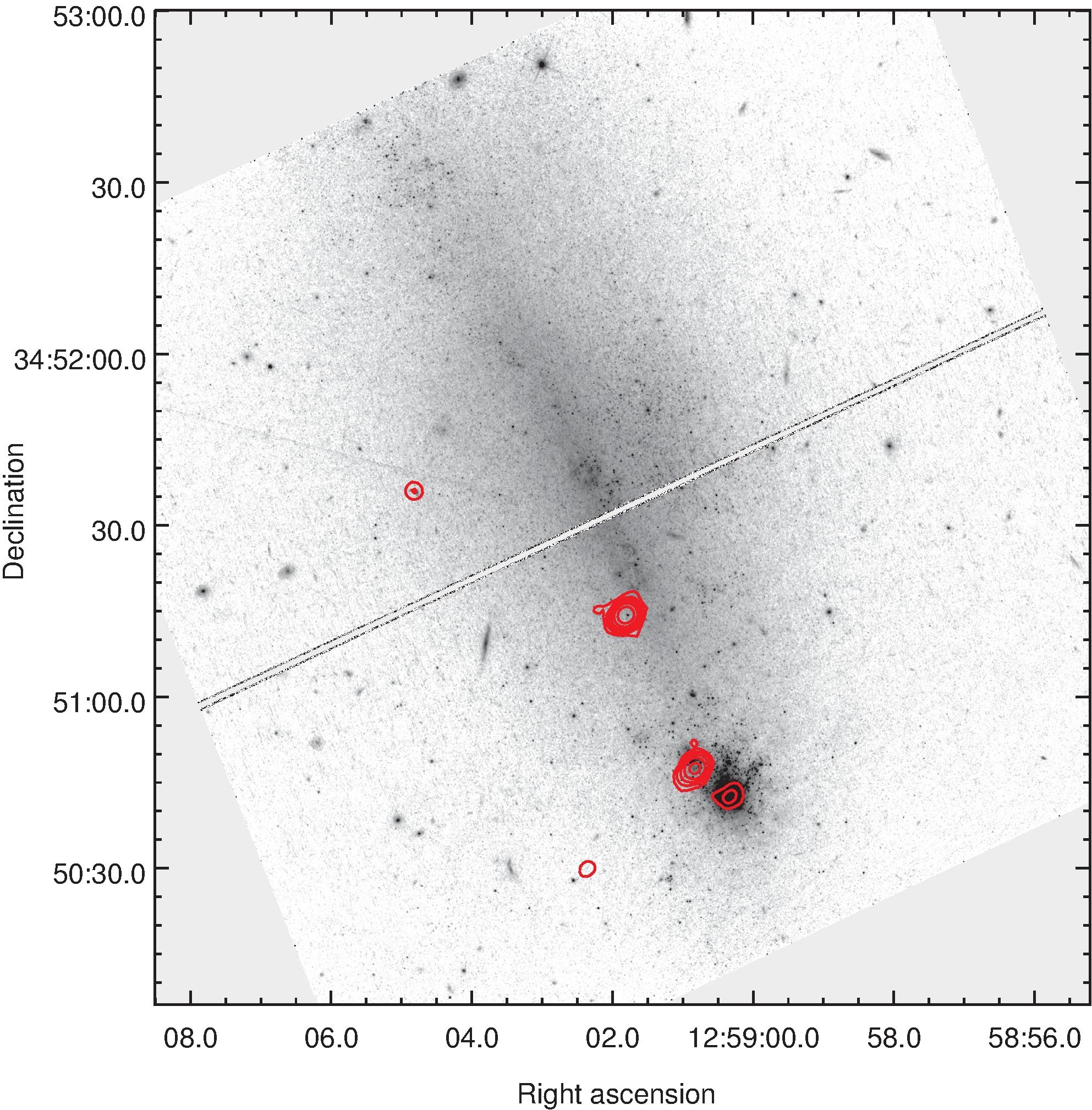}}
\hglue-0.0in{\includegraphics[height=5.3cm, angle=0]{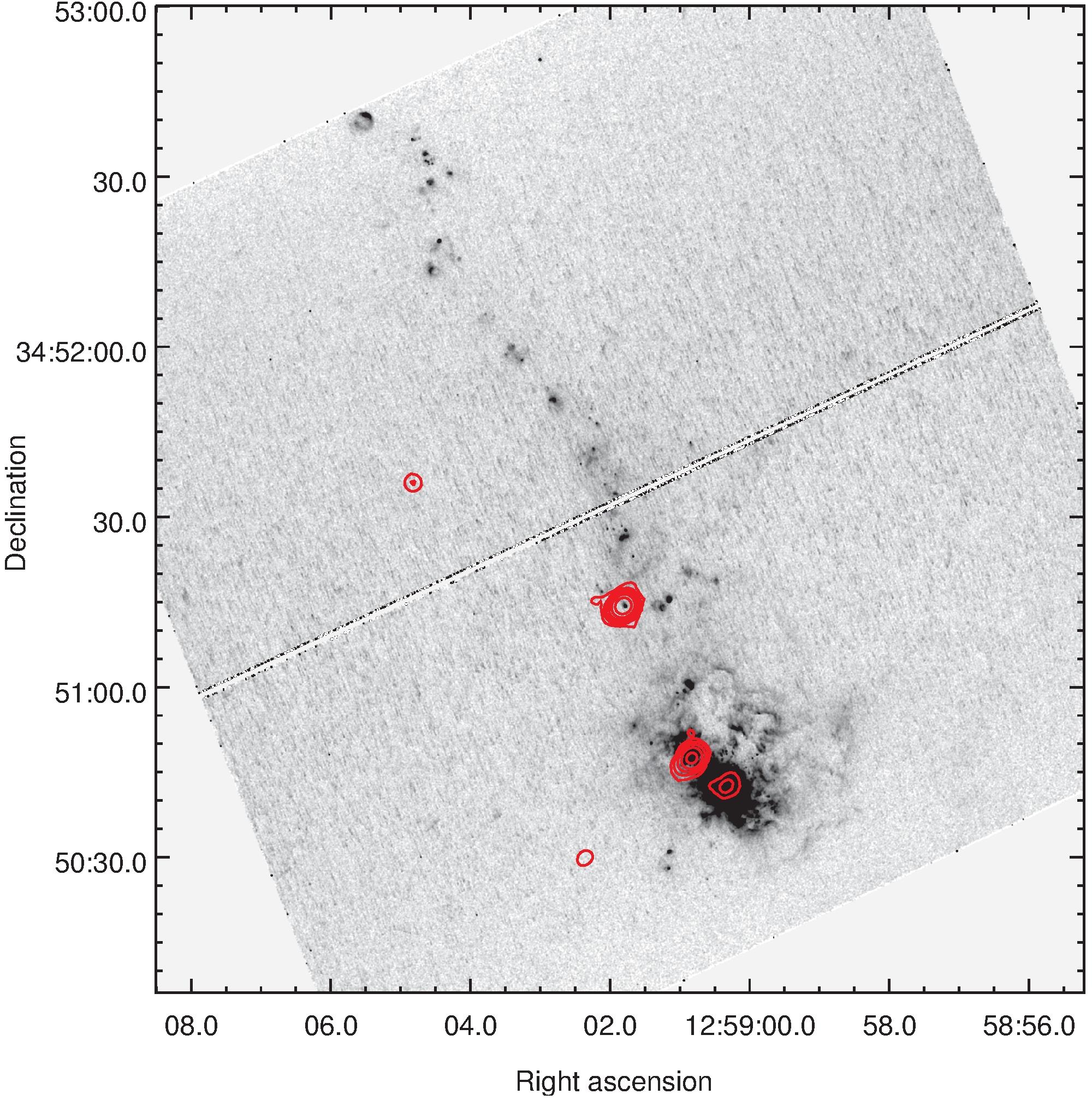}}
\end{center}
\vspace{-0.02in} \caption[Mrk59]{ X-ray contours of Mrk~59 overlaid on
  $i$-band SDSS ({\it left}), {\it HST} WFC3 F814W ({\it middle}), and
  {\it HST} WFC3 F658N line-only ({\it right})
  images. Gaussian-smoothed logarithmic contours from {\it Chandra}
  and {\it XMM-Newton} are shown in red and yellow, respectively,
  while the {\it HST} FOVs is outlined by dashed black
  lines. \label{fig:overlay1}}
\end{figure*}

\begin{figure*}
\vspace{-0.1in}
\begin{center}
\includegraphics[height=5.2cm, angle=0]{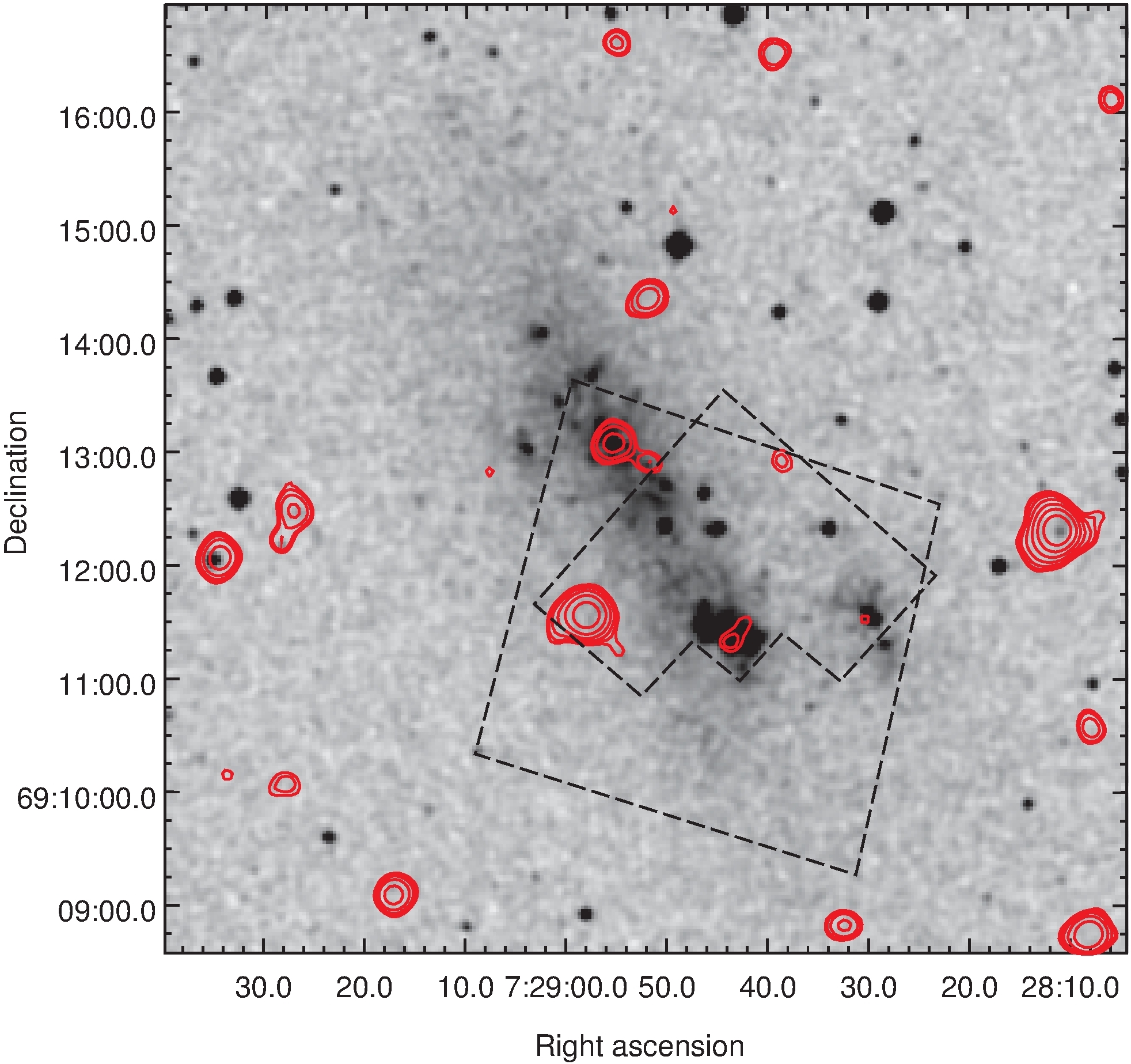}
\hglue-0.0in{\includegraphics[height=5.2cm, angle=0]{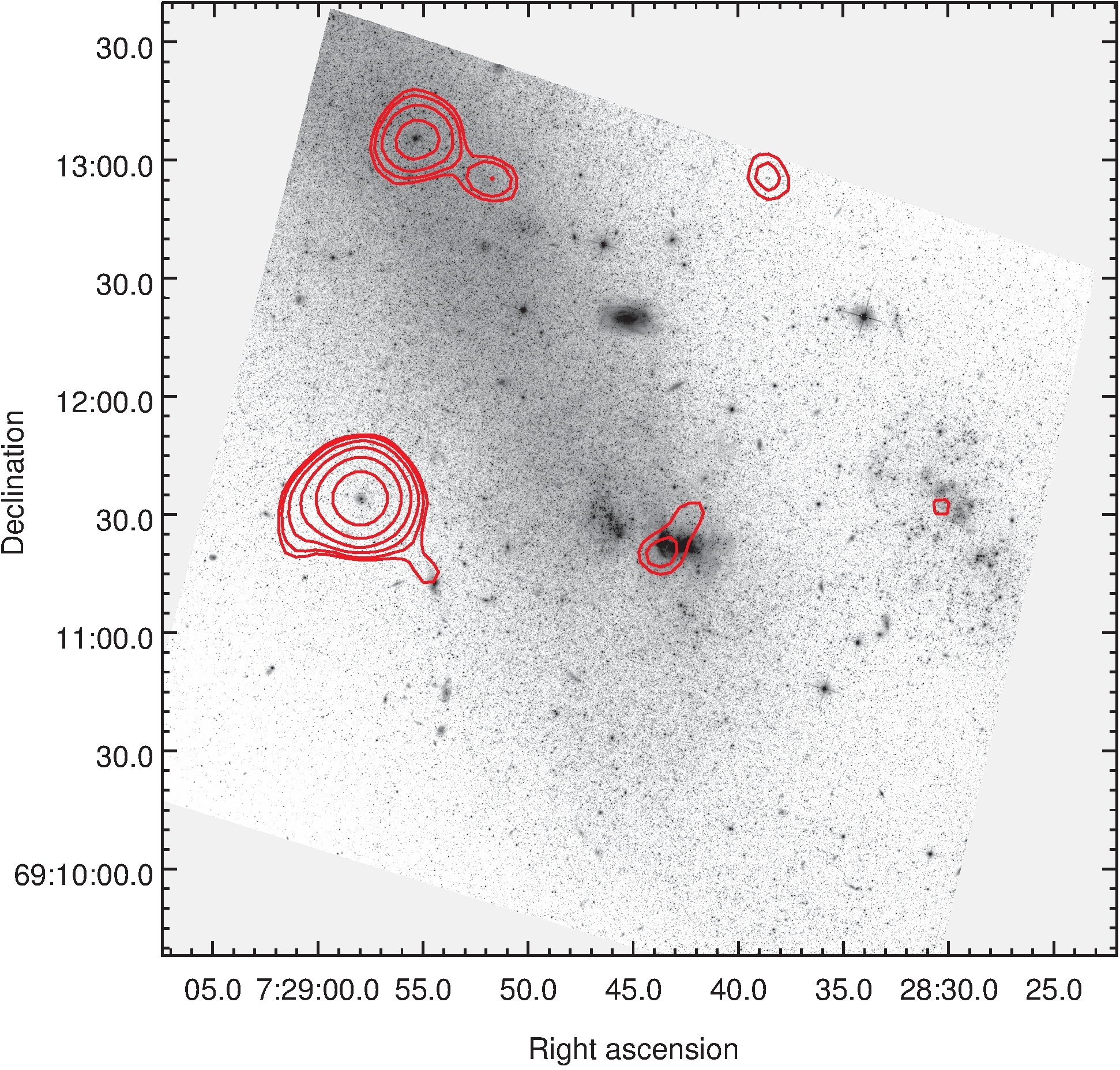}}
\hglue-0.0in{\includegraphics[height=5.25cm, angle=0]{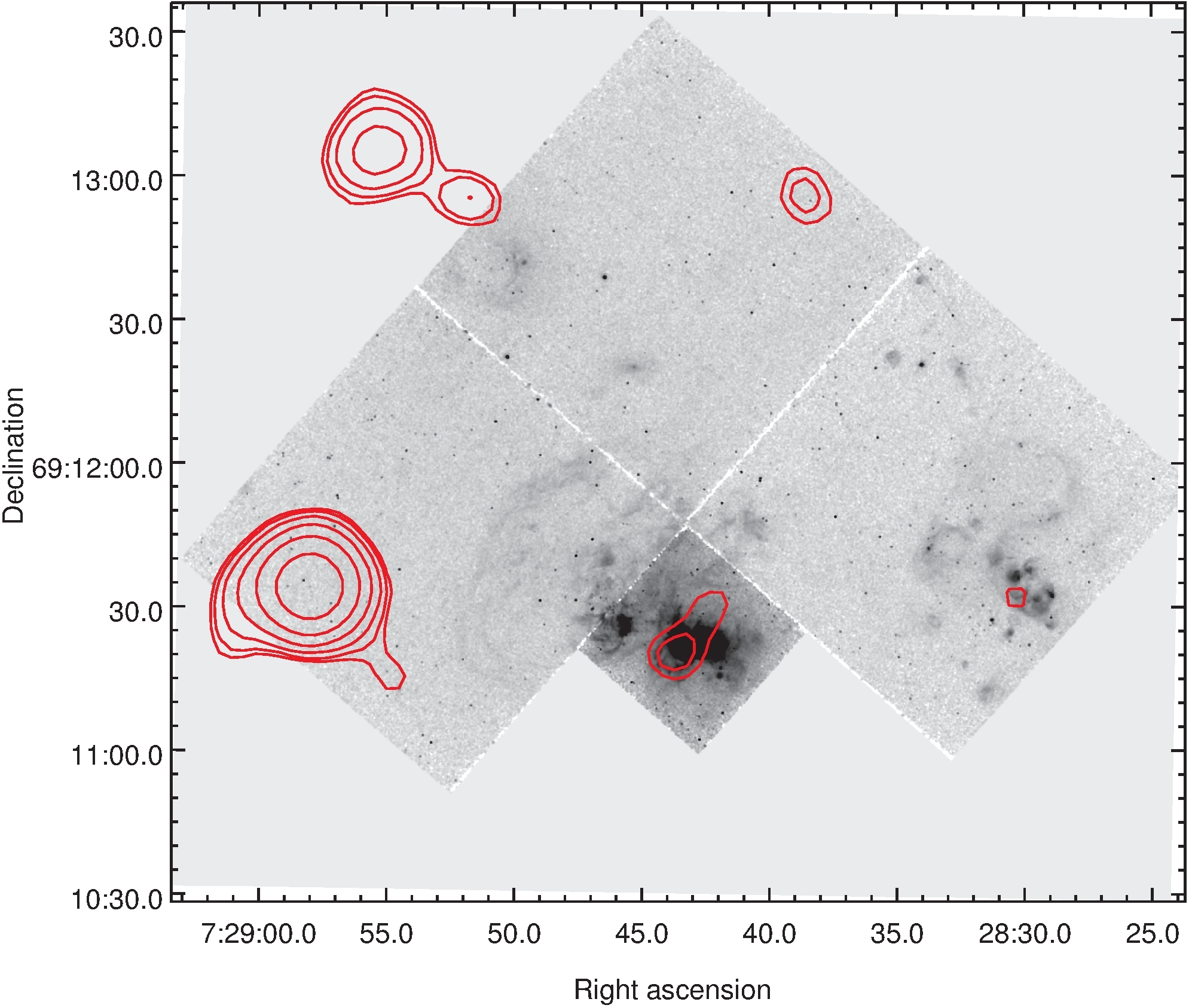}}
\end{center}
\vspace{-0.02in} \caption[Mrk71]{ X-ray contours of Mrk~71 overlaid on
  red DSS ({\it left}), {\it HST} ACS F814W ({\it middle}), and {\it
    HST} WFPC2 F656N line-only ({\it right}) images. Gaussian-smoothed
  logarithmic contours from {\it XMM-Newton} are shown in red, while
  the {\it HST} FOVs is outlined by dashed black
  lines. \label{fig:overlay2}}
\end{figure*}

\subsection{{\sl Chandra} Observations}\label{xray2}

In addition, Mrk 59 was observed by the {\sl Chandra X-ray
  Observatory} on 2012 January 3 
(PI: Yang; ObsId 12473; hereafter epoch 4, or E4) with the ACIS-I CCD
camera. The original target was the irregular galaxy NGC 4861, so
that Mrk 59 lies $\approx$2--5\arcmin off-axis from the
aimpoint. However, Chandra's point spread function (PSF) is still relatively
sharp at these off-axis angles, and its large field-of-view (FOV)
allows the entire optical extent of the galaxy 
to be imaged easily. The processing,
screening, and analysis of the data were performed using the standard
tools from {\tt CIAO} (v.4.4), as well as custom {\tt IDL}
software. No strong background flares occurred during the observation,
and the final exposure was 19.8 ks. We selected only good events for
further study.

X-ray source detection was performed on the ACIS-I image in the
0.5--8~keV band using the standard {\tt CIAO} {\it wavdetect}
algorithm with a threshold of $10^{-6}$, which conservatively
corresponds to $\la$1 false detection over the entire ACIS
detector. Six X-ray sources were found to be coincident within the
optical extent of Mrk 59 and are listed in Table~\ref{tab:Mrk59}.
The source XMMU~J125900.7+345048 was found to separate into two
sources, CXOU~J125900.87$+$345047.9 and
CXOU~J125900.37$+$345043.0. All but one of the detected sources appear
to be point-like at the resolution of {\it
  Chandra}. Background-subtracted events were extracted (and
PSF-corrected) with {\tt acis extract}, using 90\% encircled-energy
regions and local backgrounds, all of which were negligible. Our
background-subtracted detection limit corresponds to $\approx 5$
counts, or an absorbed 0.5--10~keV flux of $\approx
3.1\times10^{-15}$~erg~cm$^{-2}$~s$^{-1}$, adopting the same
assumptions as for the {\it XMM-Newton} data. Spectral products which
apply energy-dependent PSF-corrections were also generated by {\tt acis
  extract}.

\subsection{Astrometry of Optical and X-ray images}\label{optobs}

To compare the X-ray emission with the optical emission, we have
overlayed in Figures~\ref{fig:overlay1} and \ref{fig:overlay2} the X-ray contours over the following images retrieved from
various archives: for Mrk 59, an 
$i$-band Sloan Digital Sky Survey (SDSS) \citep{Abazajian2009}, {\sl
  HST} WFC3 F814W ($I$), and {\sl HST} WFC3 F658N continuum-subtracted
H$\alpha$ images; for Mrk 71, a `Red' IIIaF$+$RG610 Digital Sky Survey (DSS),
and {\sl HST} ACS F814W ($I$) and  {\sl HST} WFPC2 F656N
continuum-subtracted H$\alpha$ \citep{D00} images. The SDSS
and DSS images have a large field of view and contain enough stars for
aligning the {\sl HST} and X-ray images to the same astrometric
reference frame. To align the optical images, sources were first
extracted from the SDSS and DSS images using Sextractor
\citep[v2.2.1;][]{Bertin1996}. Optical sources from the SDSS and DSS
images were then matched to the 2MASS catalog for absolute astrometric
alignment, giving a coincidence of 32 sources within a 3$\sigma$
radius of 2\farcs2 and a rms scatter of 0\farcs61 for Mrk~59, and 93
sources within a 3$\sigma$ radius of 2\farcs1 and a rms scatter of
0\farcs58 for Mrk~71. In the same manner, the {\sl HST} images were
matched to the SDSS/DSS reference frame, yielding a coincidence of 23
sources within a 3$\sigma$ radius of 0\farcs5 and a rms scatter of
0\farcs12 for Mrk~59, and 6 sources within a 3$\sigma$ radius of
1\farcs2 and a rms scatter of 0\farcs31 for Mrk~71. Finally, the {\sl
  Chandra} and {\sl XMM-Newton} images were matched to the SDSS and
DSS frames. The {\sl Chandra} image gave a coincidence of 10
sources within a 3$\sigma$ radius of 2\farcs0 and a rms scatter of
0\farcs19 for Mrk~59, while the {\sl XMM-Newton} images yielded a
coincidence of 9 sources within a 3$\sigma$ radius of 3\farcs7 and a
rms scatter of 1\farcs03 for Mrk~59, and 13 sources within a 3$\sigma$
radius of 3\farcs5 and a rms scatter of 1\farcs0 for 
  Mrk~71.\footnote{We also tried aligning the {\sl XMM-Newton} and
    SDSS reference frames using the 3XMM-DR4 catalog positions
    \citep[e.g.,][]{Watson2009}; (Watson et al. in preparation) as the 3XMM-DR4 catalog
    provides some astrometric corrections and gains over the {\sl
      XMM-Newton} data alone. These positions further improved the overall 
    alignment for the majority of the XMM sources to a rms scatter of
    $\approx$0.\farcs6 for Mrk~59, but oddly yielded systematic
    offsets of 1\farcs5-2\farcs0 to the southwest for the two 
brightest sources IXO 72 and 73,
    compared to the positions we calculated from our own
    astrometrically-corrected {\sl Chandra} and {\sl XMM-Newton}
    images.  Thus we use our own aligned positions throughout this work.}
  Overall, our optical alignment improved substantially upon the
  quoted baseline {\sl Chandra} and {\sl XMM-Newton} astrometric
  accuracies of \hbox{$\approx$0\farcs6} \footnote{see
    http://cxc.harvard.edu/cal/ASPECT/celmon} and
  \hbox{$\approx2--4$\arcsec} \citep[e.g.,][]{Jansen2001, Watson2009},
  respectively.

\begin{figure*}
\vspace{-0.1in}
\begin{center}
\includegraphics[width=18cm, angle=0]{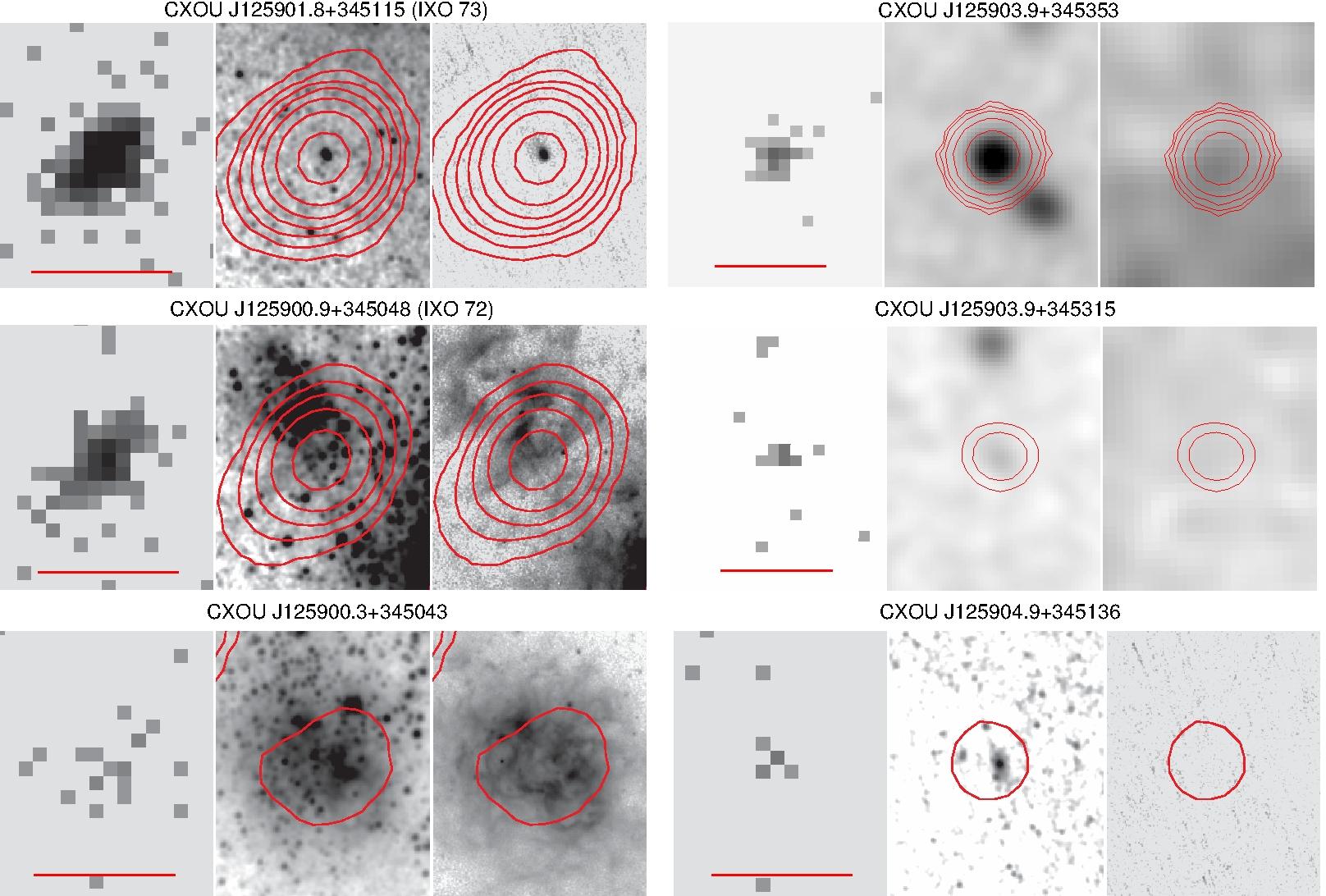}
\end{center}
\vspace{-0.02in} \caption[Mrk59b]{Cutout images of the X-ray-detected
  sources which lie within the optical extent of Mrk~59. Three-panel
  cutouts show {\it Chandra} ({\it left}), {\it HST} WFC3 F814W or
  {\it SDSS} i-band ({\it middle}), and {\it HST} WFC3 F658N or
  Palomar H$\alpha$ \citet{GildePaz2003} line-only
  images ({\it right}). Gaussian-smoothed logarithmic contours from
  {\it Chandra} are shown in red. The red line denotes an angular size
  of 5\arcsec{}. In all cases, North is up and East is left. \label{fig:Mrk59_cutouts}}
\end{figure*}

\begin{figure}
\vspace{-0.1in}
\begin{center}
\includegraphics[width=9cm, angle=0]{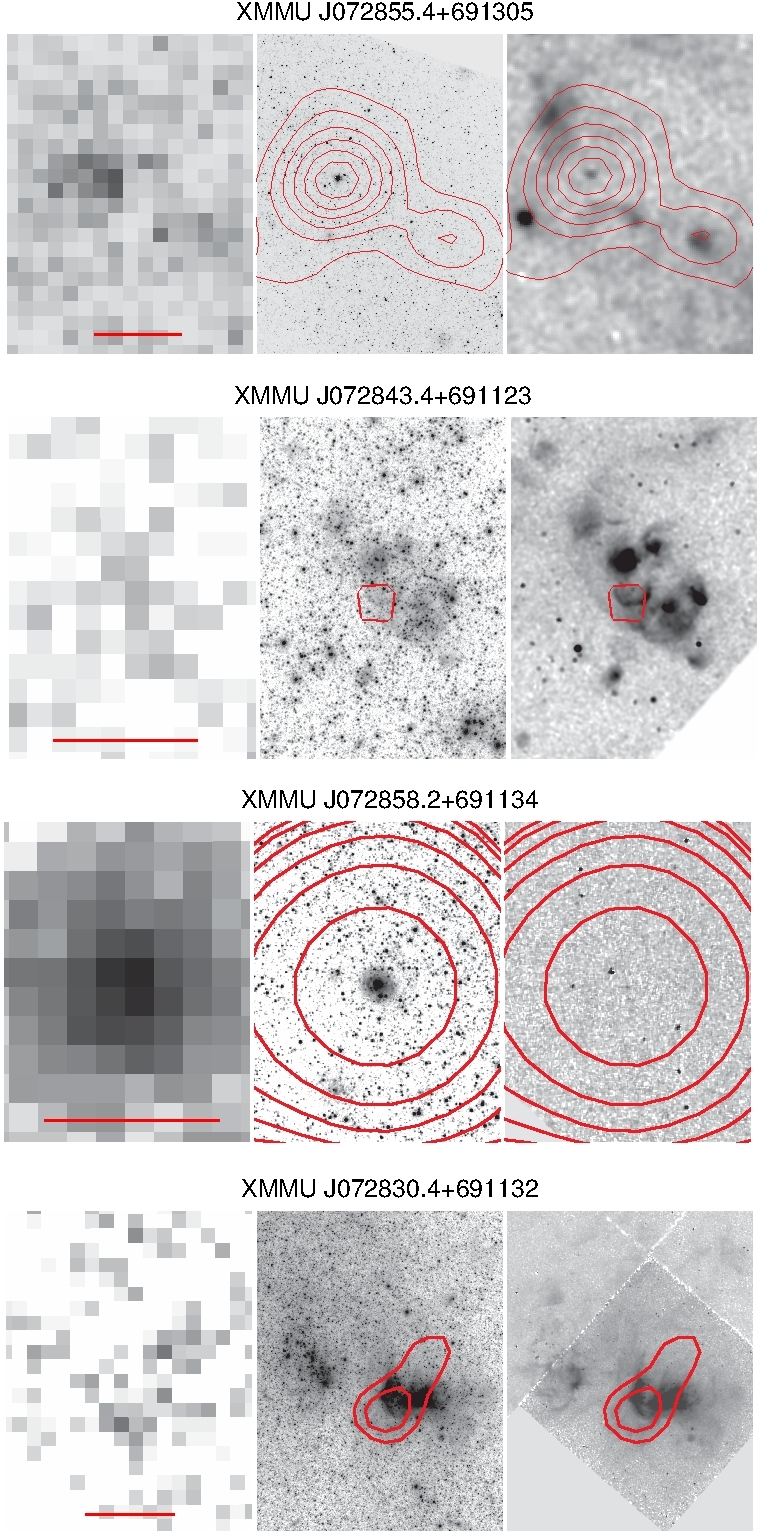}
\end{center}
\vspace{-0.02in} \caption[Mrk59b]{Cutout images of the X-ray-detected
  sources which lie within the optical extent of Mrk~71. Three-panel
  cutouts show {\it XMM} ({\it left}), {\it HST} ACS F814W ({\it
    middle}), and {\it HST} WFPC2 F656N or Kapteyn H$\alpha$
  \citep{James2004} line-only images ({\it right}). Gaussian-smoothed
  logarithmic contours from {\it XMM-Newton} are shown in red. The red
  line denotes an angular size of
  15\arcsec{}. In all cases, North is up and East is left. \label{fig:Mrk71_cutouts}}
\end{figure}

\subsection{X-ray Counterparts}\label{xcntrparts}

From the X-ray contour plots in Figures~\ref{fig:overlay1} and
\ref{fig:overlay2}, it appears that Mrk~59 consists of two very bright
X-ray point sources (CXOU~J125901.85$+$345114.7 and CXOU~J125900.87$+$345047.9) 
and two possible faint ones 
(CXOU~J125904.85$+$345136.6 and CXOU~J125903.95$+$345315.3). The
X-ray point sources in Mrk~59 are located along the
cometary tail, extending from the main H~{\sc ii} region which itself
exhibits slight hints of diffuse X-ray emission
(CXOU~J125900.37$+$345043.0). On the other hand, 
Mrk~71 has only a few relatively
faint X-ray point sources (XMMU~J072857.9+691135 and
XMMU~J072855.2+691305) in addition to traces of diffuse emission
associated with its numerous H~{\sc ii} regions (most notably
XMMU~J072843.0+691123 and XMMU~J072830.4+691134). The likelihood of
chance superpositions with Galactic X-ray sources and background AGN
within the optical extent of either galaxy is small, but
non-negligible.  From extrapolation of the medium-deep {\sl
  XMM-Newton} $\log N$--$\log S$ relation \citep[e.g.,][]{Baldi2002} to their
respective flux limits, we would expect on average $\sim$~1 foreground/background
source within the optical extent of either Mrk~59 and Mrk~71.

Given the high optical source densities in both galaxies and the
relative uncertainty in the centroids of the {\sl XMM-Newton} sources,
identifying optical counterparts of {\sl XMM-Newton} detections with
any reasonable degree of certainty is difficult. Identifying
counterparts of {\sl Chandra} sources, however, is substantially more secure.
All of the counterpart identifications are shown in
Figures~\ref{fig:Mrk59_cutouts} and \ref{fig:Mrk71_cutouts} and
described in $\S$\ref{individual}.

When we consider an X-ray source to be associated with 
Mrk~59 or Mrk~71, we also
calculate its X-ray luminosity accordingly.

\subsection{X-ray Properties}\label{xprops}

The basic X-ray properties of the sources are listed in 
  Table~\ref{tab:Mrk59} for Mrk 59 and Table~\ref{tab:Mrk71} for Mrk 71. 
By design, all of the
  sources are detected either in the {\sl Chandra} 0.5--8.0~keV band or
  the {\sl XMM-Newton} 0.5--10.0~keV band. The vast majority of these
  detections are seen in the soft band (0.5-2.0 keV), while only a
  minority are detected in the hard band (2-8 keV). Thus,
  the observed hardness ratios (the definition of which is 
given in the notes to Table~\ref{tab:Mrk59})  
indicate that the majority of the
  detected objects have predominantly soft spectral characteristics, 
as expected. Only
  CXOU~J125904.9$+$345136 and XMMU~J072857.9$+$691135 exhibit 
  relatively hard spectra. For sources with more than
  100 counts, we compute the X-ray fluxes and unabsorbed luminosities
  from direct spectral fitting. For those below this limit, we
  calculate these quantities assuming an absorbed power-law model ({\tt
    tbabs*powerlaw}) with $\Gamma$$=2$ and Galactic $N_{\rm H}$ for
  low hardness-ratio (HR) sources (HR$\le$1), and $N_{\rm
    H}=3\times10^{21}$ for high hardness-ratio sources (HR$>$1). Note
  that if the spectrum of a faint source deviates substantially from
  these average values, then the flux and absorption-corrected
  luminosity of the source may change as well.

The {\it Chandra} and {\it XMM-Newton} spectra for the bright sources
were analyzed using {\tt XSPEC} \citep{Arnaud1996}. Unless stated
otherwise, spectral parameter errors are for 90\% confidence, assuming
one parameter of interest ($\Delta\chi^2=2.7$). None of the sources
was affected by pile-up. We fitted the spectra in {\tt XSPEC} with
simple spectral models. The intrinsic $N_{\rm H, intr}$, spectral
slopes, and normalizations were allowed to vary, while the Galactic
column densities and redshifts were fixed at their fiducial
values. The X-ray spectra and their best-fit models are shown in
Figures~\ref{fig:Mrk59_IXO72_spectra}, \ref{fig:Mrk59_IXO73_spectra}, and \ref{fig:Mrk71_X1_spectra}. The X-ray fluxes and
absorption-corrected luminosities were calculated using these simple
models with {\tt XSPEC}. We grouped the spectra to have at least 15
counts per bin and employed $\chi^2$ statistics.

We discuss spectral fitting results and hardness ratios for individual
sources in $\S$\ref{individual}.

\subsection{X-ray Timing Analysis}\label{timing}

Although the {\sl XMM-Newton} and {\sl Chandra} observations of Mrk~59
were relatively short, we do have adequate statistics to evaluate the
short-term timing characteristics for the two bright point
sources. Mrk~59 was additionally observed with the {\sl ROSAT} HRI, so
we also can assess its long-term soft-band variability as
well. Unfortunately, Mrk~71 was not observed by any previous X-ray
observatory and the statistics for the faint sources in this galaxy
provide no useful constraints; thus we only discuss Mrk~59
hereafter. To examine objectively the existence of any significant
variation in the count rate for the two bright X-ray sources in
Mrk~59, we used the Kolmogorov-Smirnov (KS) statistic on the unbinned
data to test the null hypothesis that the count rate for 
each source plus the background rate
was constant over the duration of individual exposures. Over the short
timescales of our observations ($\la$8 hr), neither of the bright
sources in Mrk~59 varied significantly at the $>$~90\% confidence
level.

To constrain longer-term variations for the two bright X-ray
  point sources in Mrk 59, we compare the {\sl XMM-Newton} and {\sl
  Chandra} observations in the 0.5--10.0 keV flux range. This provides 
a 10 yr baseline. We also compare these data with  
the 21.4~ks {\sl ROSAT} HRI observation on 1992 June 24--27,
which provides a $\approx$20~yr baseline.  To this end, we extracted
background-subtracted HRI counts at the position of each source, using
apertures of 15$\arcsec$ and local backgrounds. Because the {\sl ROSAT} data do not cover the same band (only the 0.1--2.4 keV range), 
and do not have good spectral constraints, we can only really compare the 
0.5-2.0 keV fluxes which we determined by using best-fitted models to the   
{\sl XMM-Newton} and {\sl Chandra} observations. 
We discuss long-term
variability results for individual 
sources in the next section.


\section{Individual X-ray sources}\label{individual}

\subsection{Mrk 59}


\subsubsection{CXOU~J125901.8$+$345115 (IXO 73)}\label{59_X1}

CXOU~J125901.8$+$345115, also known as IXO 73
\citep{Colbert2002}, is the brightest source in Mrk~59. Its X-ray properties 
have already been discussed in several studies of ultraluminous X-ray
sources (ULXs), based on the same {\it XMM-Newton} data analyzed here
\citep[e.g.,][]{Colbert2002, Liu2005a, Liu2005b,
  Lopez-Corredoira2006, Stobbart2006, Heil2009, Yang2012}. 

Comparing the {\it Chandra} position of IXO 73 with the {\it HST} WFC3
data in Figure~\ref{fig:Mrk59_cutouts}, we see that the X-ray source
is coincident with a moderately isolated bright point
source with $m_{\rm F814W}=21.7$ ($m_{\rm F547M}=21.8$, $m_{\rm F658N,
  line-only}=18.2$, $m_{\rm F606W}=21.5$, $m_{\rm F487N}=20.3$), lying
$\approx$0\farcs1 to the northwest.  
At the distance of Mrk~59, this
corresponds to an uncorrected absolute magnitude of $M_{\rm 
F814W}=-8.5$, comparable to the intrinsic brightness of
the brightest late-type supergiant O stars or hypergiant stars.
From the narrow-band F658N image,
it can be seen that IXO 73 
is associated with a luminous 
H {\sc ii} region 
in Mrk~59.
\citet{Stobbart2006} performed spectral fitting for IXO 73 using the
{\sl XMM-Newton} data, and showed that its spectrum can be  
fitted by a number of simple and physically-motivated models. We
carried out joint spectral fitting of the {\sl XMM-Newton} and {\sl
  Chandra} data for IXO 73, adopting one of their simple models, the one 
which includes a power-law continuum plus a cool disk blackbody [{\tt
    tbabs*ztbabs*(pow+diskbb)}], so as to provide a point of
comparison with the extensive \citet{Stobbart2006} work. We fixed the
Galactic neutral hydrogen column density to its fiducial value and
allowed the intrinsic column density and two-component temperatures
and normalizations to vary. We initially adopted a simple model
wherein all epochs could be characterized by a single set of
parameters.  Only the overall normalizations were allowed to vary by a
constant value to account for calibration uncertainties and/or flux
variations. With this model, all epochs were acceptably fitted by the
following parameter set: 
$N_{\rm H, intr}=1.00^{+0.11}_{-0.08}\times10^{21}$~cm$^{-2}$,
$\Gamma_{pow}=2.17^{+0.03}_{-0.03}$,
$N_{pow}=5.91^{+0.29}_{-0.21}\times10^{-5}$ photons\,cm$^{-2}$\,s$^{-1}$\,keV$^{-1}$ at 1 keV,
$kT_{disk}=0.21^{+0.01}_{-0.02}$~keV, 
$N_{disk}=2.03^{+1.42}{-0.70}$ [in units of (($R_{in}$/km)/(D/10 kpc))$^{2}$ $\cos{\Theta}$], and
relative normalization factors of 1.00 (fixed),
0.79$^{+0.01}_{-0.01}$, 
0.61$^{+0.02}_{-0.02}$ and
2.28$^{+0.03}_{-0.03}$ 
for epochs E1 to E4, with a $\chi^{2}=221.49$
for 235 degrees of freedom. If we decouple the model parameters for
each epoch, and let all temperature and normalization components vary
separately, we find mild shifts in the best-fitted temperatures and
normalizations, with a $\chi^{2}=203.30$ for 218 degrees of
freedom. The errors on the parameter values for each epoch are large,
however, such that the values completely overlapped amongst the epochs
within their respective $3\sigma$ errors. The improvement over the
fixed parameter fit was only at the $\approx 1.8\sigma$ level,
according to the f-test. Thus, we adopt the simpler fixed model. These
simple model parameter values are in good agreement with those found by
\citet{Stobbart2006}. The large changes in renormalization factors
between different epochs demonstrate that IXO 73 has varied by at
least a factor of $\approx$3.7 over a $\approx$10 yr span. Such strong
variability is not uncommon for ULXs. The observed 0.5--10.0 keV
fluxes and unabsorbed luminosities are given in Table~\ref{tab:Mrk59},
while the individual spectra for each epoch are shown in
Figure~\ref{fig:Mrk59_IXO73_spectra}. 

Using the same spectral model as above, we obtained a soft-band {\sl
  ROSAT} HRI flux of $F_{\rm 0.5-2~keV}\approx
1.8\times10^{-13}$~erg~cm$^{-2}$~s$^{-1}$ for IXO 73. Comparing this
to the 0.5--2.0 keV fluxes of $1.16\times10^{-13}$~erg~cm$^{-2}$~s$^{-1}$
from {\sl XMM-Newton} and $2.5\times10^{-13}$~erg~cm$^{-2}$~s$^{-1}$
from {\sl Chandra}, we find that IXO 73 has experienced 
a soft-band long-term variability of a factor of $\sim$2.

\subsubsection{CXOU~J125900.9$+$345048 (IXO 72)}\label{59_X2}

CXOU~J125900.9$+$345048, also known as IXO 72
\citep{Colbert2002}, is the second brightest source in Mrk~59. It too has
been previously well-characterized in several studies of ULXs, using the
{\it XMM-Newton} data \citep[e.g.,][]{Colbert2002, Liu2005b,
  Lopez-Corredoira2006, Sanchez-Sutil2006, Yang2012}.

Overlaying the {\it Chandra} X-ray contours of IXO 72 on the {\it HST} WFC3
images in Figure~\ref{fig:Mrk59_cutouts}, we see that the X-ray source
is located in a dense stellar cluster and is best-matched by a
relatively isolated point source with $m_{\rm F814W}=23.0$ ($m_{\rm
  F547M}=22.7$, $m_{\rm F658N, line-only}=19.4$, $m_{\rm F487N}=21.8$)
which lies 0\farcs1 to the northwest. At the distance of Mrk~59, this
corresponds to an uncorrected absolute magnitude of $M_{\rm
  F814W}=-7.1$, comparable to the intrinsic brightness of
a late-type supergiant O star. This identification is however not unique: 
there are a few other objects several magnitudes fainter
($m_{\rm F814W}\ga24.3$) which could also be matches to the X-ray
source within the positional error. Thus, some caution should be
exercised when identifying the X-ray source. IXO 72 does resides however on the
outskirts of a massive H {\sc ii} complex and lies on top of a
significant knot of H$\alpha$ emission coincident with the location of
the brightest optical candidate counterpart. This implies that the X-ray
source and its presumed optical counterpart are most likely   
associated with this H {\sc ii} region complex.
The spectral
observation may include all (both point-like and diffuse) emission
within 1--2$\arcsec$ of the true counterpart of IXO 72. Thus, the {\sl
  XMM-Newton} observations of IXO 72 may contain some slight
contamination from CXOU~J125900.3$+$345043. However, given the flux
difference between the 2 sources 
and the consistency of the fluxes of IXO 72 between the E3 and E4  
epochs (the latter being based on {\it Chandra} data), we believe 
that contamination should not greatly 
affect the {\sl XMM-Newton} results,
unless CXOU~J125900.3$+$345043 varied dramatically.

\begin{figure*}
\vspace{0.1in}
\centerline{
\includegraphics[height=9cm, angle=-90]{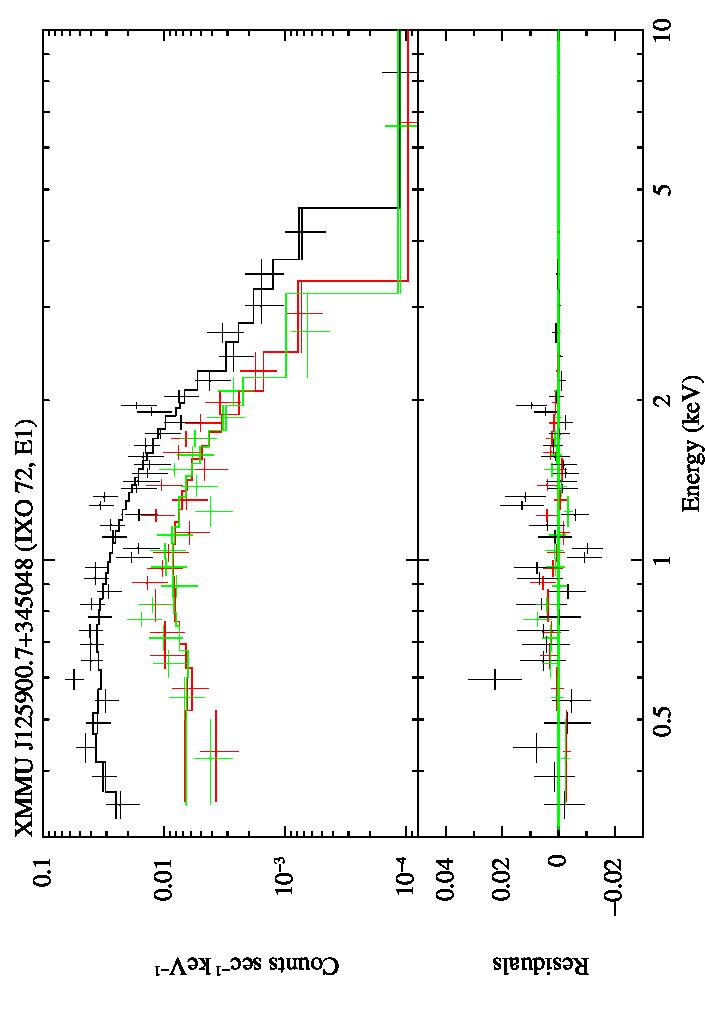}
\hfill
\includegraphics[height=9cm, angle=-90]{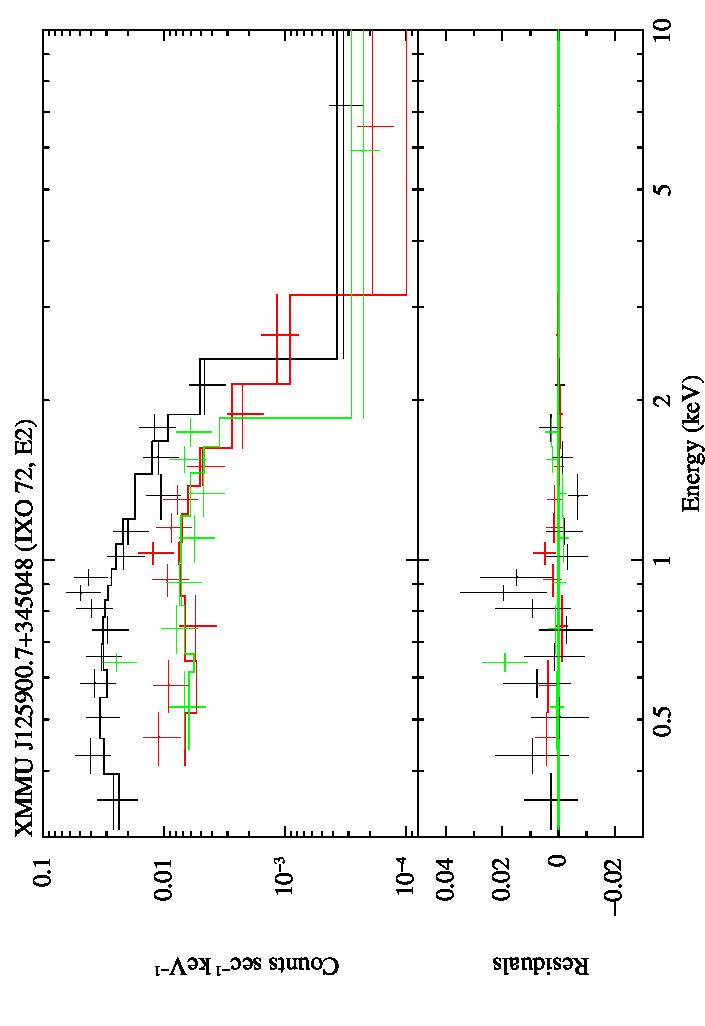}
}
\centerline{
\includegraphics[height=9cm, angle=-90]{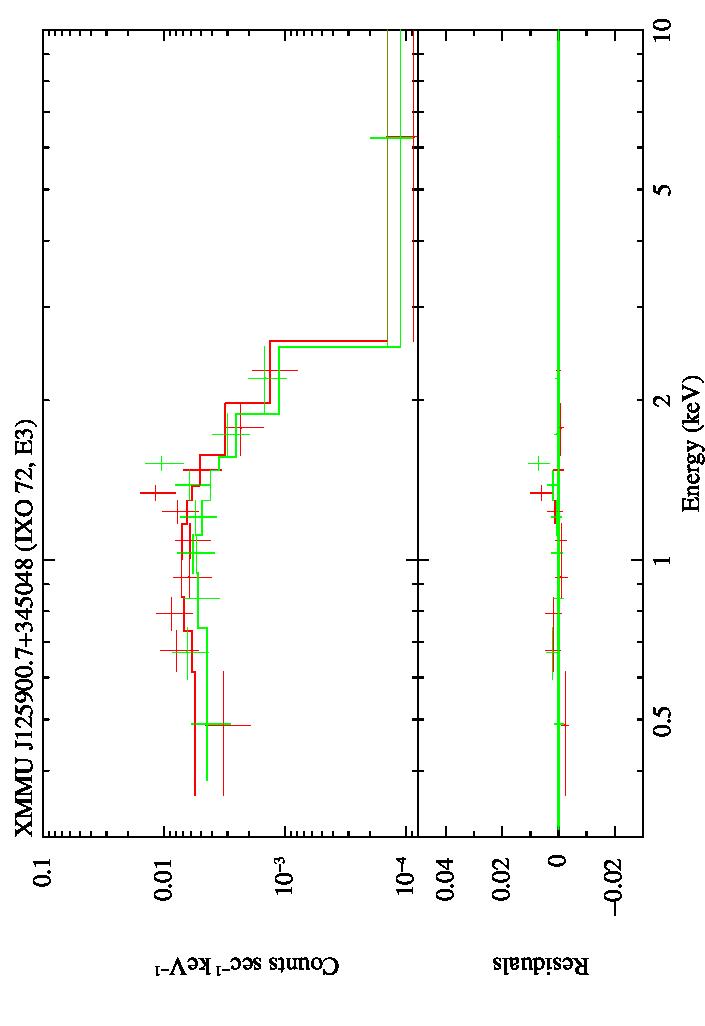}
\hfill
\includegraphics[height=9cm, angle=-90]{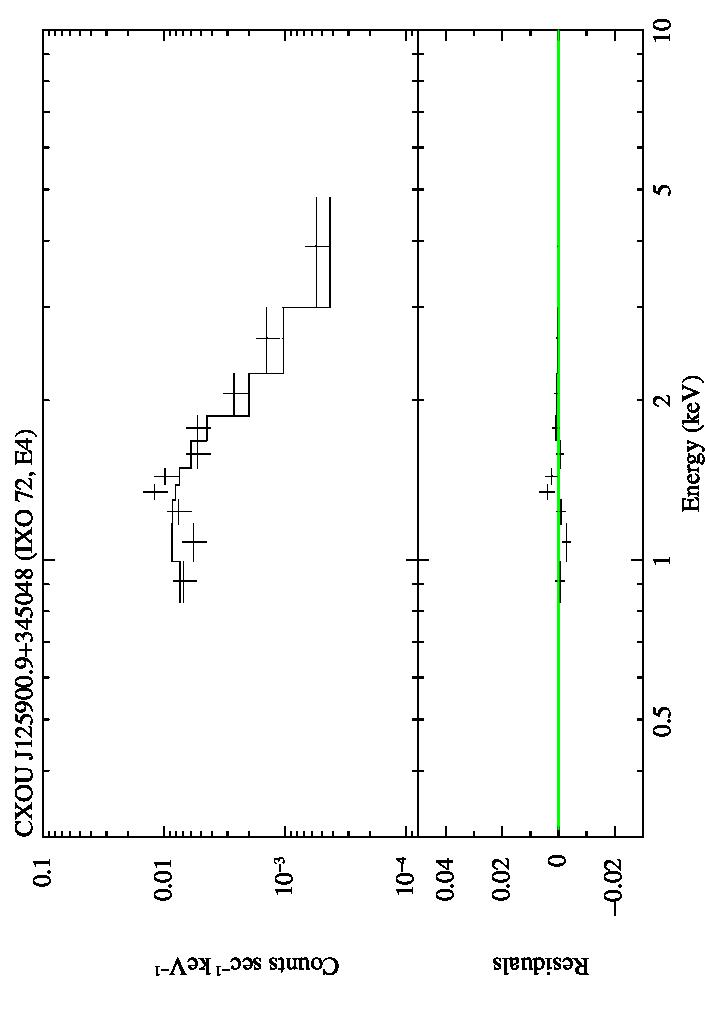}
}
\vspace{0.1cm} 
\caption[]{Four epochs of X-ray spectra of IXO 72 in Mrk 59. The first three epochs show {\sl XMM-Newton}
  $pn$, MOS1, and MOS2 spectra as black, red and green, respectively,
  while the last epoch shows the {\sl Chandra} ACIS-I spectra in
  black.  The upper panel of each plot presents the X-ray spectra and
  best-fit models (see $\S$\ref{individual} for details), while the lower
  panels show the residuals of the fits for the best-fit
  model.\label{fig:Mrk59_IXO72_spectra}}
\vspace{0.2cm} 
\end{figure*}

\begin{figure*}
\vspace{0.1in}
\centerline{
\includegraphics[height=9cm, angle=-90]{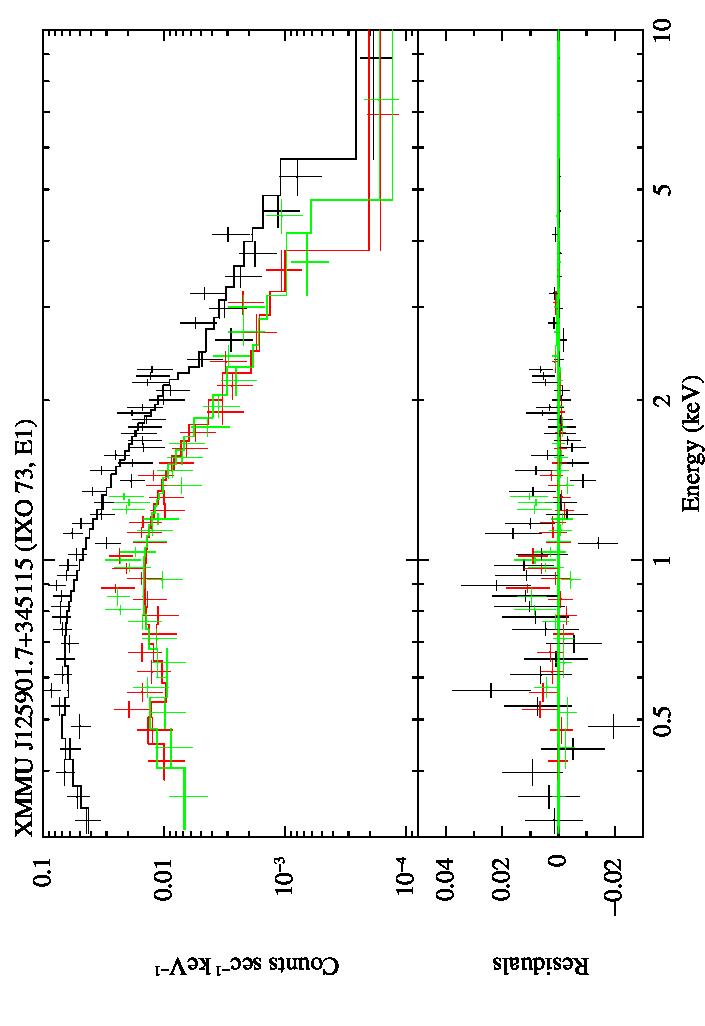}
\hfill
\includegraphics[height=9cm, angle=-90]{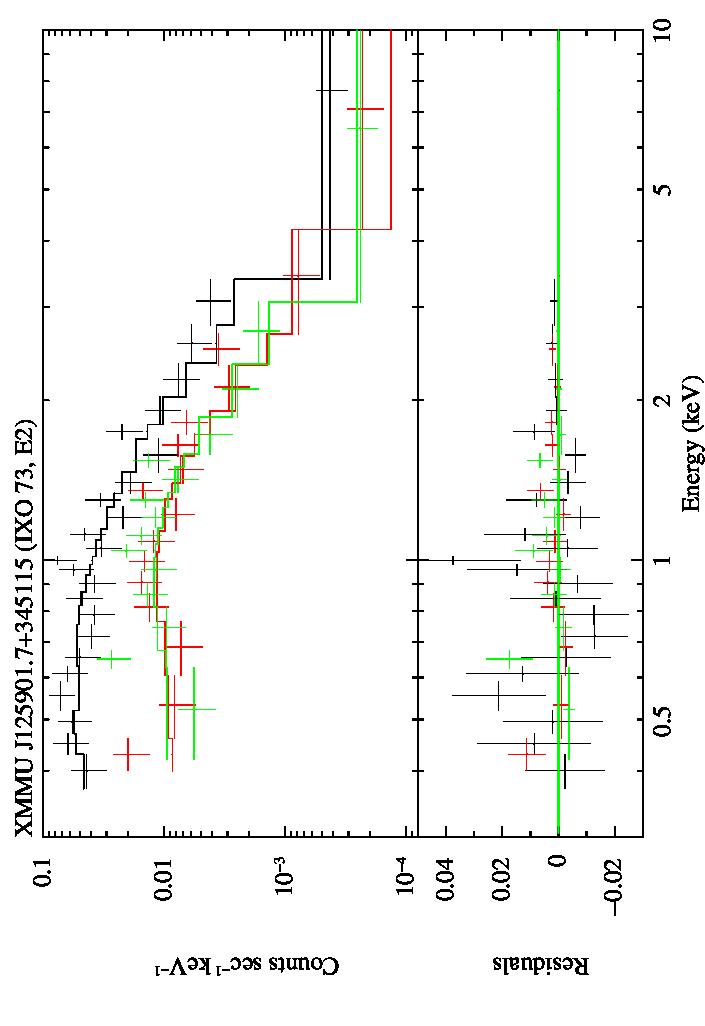}
}
\centerline{
\includegraphics[height=9cm, angle=-90]{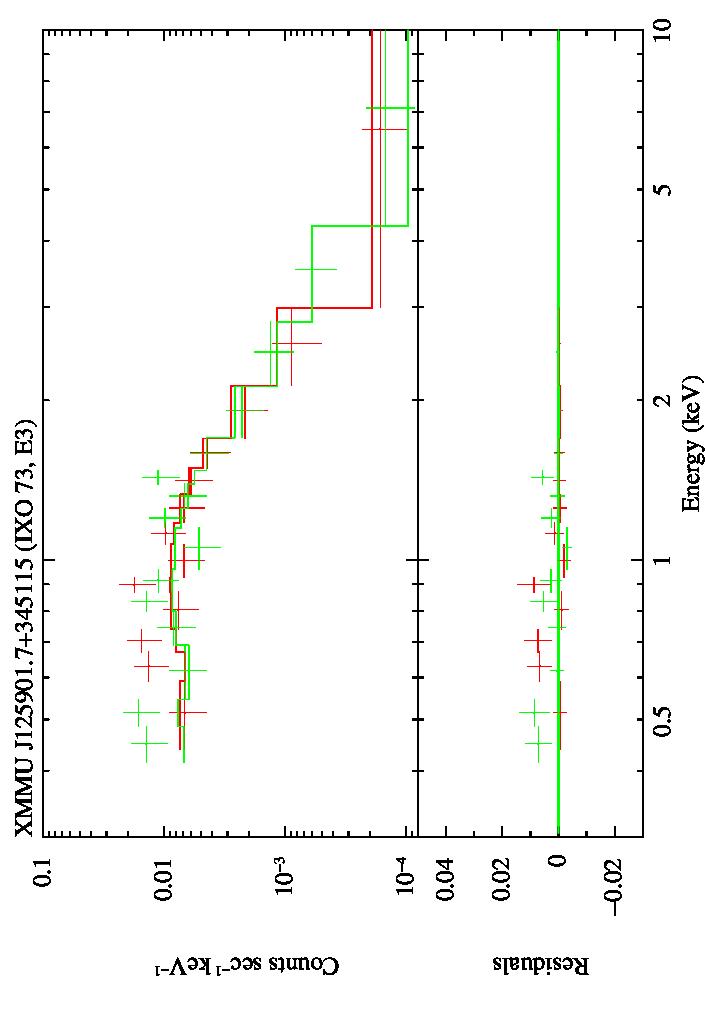}
\hfill
\includegraphics[height=9cm, angle=-90]{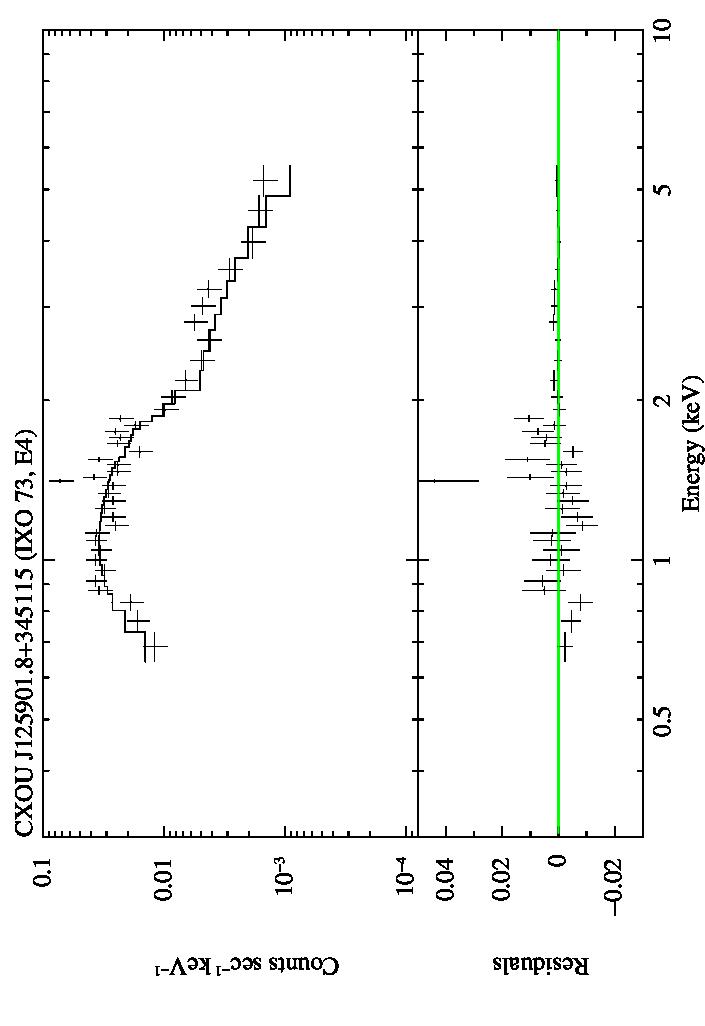}
}
\vspace{0.1cm} 
\caption[]{Four epochs of X-ray spectra of IXO 73 in Mrk 59. 
The first three epochs show {\sl XMM-Newton}
  $pn$, MOS1, and MOS2 spectra as black, red and green, respectively,
  while the last epoch shows the {\sl Chandra} ACIS-I spectra in
  black.  The upper panel of each plot presents the X-ray spectra and
  best-fit models (see $\S$\ref{individual} for details), while the lower
  panels show the residuals of the fits for the best-fit
  model.\label{fig:Mrk59_IXO73_spectra}}
\vspace{0.2cm} 
\end{figure*}

We carried out a joint spectral fitting of the {\sl XMM-Newton} and
{\sl Chandra} data for IXO 72, again adopting a simple model, [{\tt
    tbabs*ztbabs*(pow+diskbb)}], to facilitate comparison to IXO 73 and
the work of \citet{Stobbart2006}.  As with IXO 73, we fixed the
Galactic neutral hydrogen column density to its fiducial value, and
allowed the intrinsic column density, two-component temperatures and
normalizations to vary. We initially adopted the simple model wherein
all epochs can be described by a single parameter set, with only the
overall normalizations allowed to vary by a constant value.  All
epochs were acceptably fitted by this model with parameter values of
$N_{\rm H, intr}=1.3^{+0.2}_{-0.1}\times10^{21}$~cm$^{-2}$,
$\Gamma_{pow}=2.72^{+0.14}_{-0.11}$,
$N_{pow}=3.88^{+0.29}_{-0.33}\times10^{-5}$ photons\,cm$^{-2}$\,s$^{-1}$\,keV$^{-1}$ at 1 keV,
$T_{disk}=0.83^{+0.10}_{-0.10}$~keV, 
$N_{disk}=0.003^{+0.002}{-0.001}$ [in units of (($R_{in}$/km)/(D/10 kpc))$^{2}$ $\cos{\Theta}$], and
and normalization factors of 1.00 (fixed), 
$0.89^{+0.02}_{-0.02}$,
$0.86^{+0.02}_{-0.02}$, and 
$0.97^{+0.03}_{-0.02}$ 
for the different epochs, with a $\chi^{2}=133.29$ for 137 degrees of
freedom.  If we decouple the model parameters for the {\sl XMM-Newton}
and {\sl Chandra} spectra and fit the data again, we see mild shifts
in the best-fitted values and a $\approx 1.3\sigma$ improvement in
$\chi^{2}=119.55$ for 128 degrees of freedom according to the f-test,
although the values completely overlap amongst the epochs within their
respective $3\sigma$ errors.   \citet{Stobbart2006} 
did not include IXO 72 in their analysis, but 
these simple model parameter values are
in general agreement with those found by them for
other ULXs.  The {\sl XMM-Newton} renormalization factors for epochs
E2 and E3 imply the source has undergone only mild variability
($\approx$1.1) over a $\approx$10 yr span. The observed 0.5--10.0 keV
fluxes and unabsorbed luminosities are given in Table~\ref{tab:Mrk59},
while the individual spectra for each epoch are shown in
Figure~\ref{fig:Mrk59_IXO72_spectra}.

Using the same spectral model as above, we obtained a soft-band {\sl
  ROSAT} HRI flux of $F_{\rm 0.5-2~keV}\approx
6.7\times10^{-14}$~erg~cm$^{-2}$~s$^{-1}$ for IXO 72. Comparing this
to the 0.5--2.0 keV fluxes of $6.59\times10^{-14}$~erg~cm$^{-2}$~s$^{-1}$
from {\sl XMM-Newton} and $6.57\times10^{-14}$~erg~cm$^{-2}$~s$^{-1}$
from {\sl Chandra}, the soft band fluxes of 
IXO 72 appears to have remained constant over a
20~yr timespan within the errors.

\subsubsection{CXOU~J125900.3$+$345043}\label{59_X3}

CXOU~J125900.3$+$345043 is seen by {\sl Chandra} as 
  distinct from IXO 72, but the two sources are not separate at the
  resolution of {\sl XMM-Newton}. This source may be extended, as its
  radial profile appears marginally broader (a $\approx$2$\sigma$
  deviation based on Cash statistics; Cash 1979) as compared to the {\sl
    Chandra} PSF. Comparing the {\it Chandra} position with the {\it
    HST} WFC3 data in Figure~\ref{fig:Mrk59_cutouts}, we see that the
  X-ray source aligns well with a dense stellar cluster and massive
  H {\sc ii} complex at the ``head'' of Mrk~59 \citep[see
    also][]{Yang2012}. The hardness ratio of the source indicates it
  is relatively soft, but given the low number of counts, it remains
  unclear whether the X-ray emission arises from hot gas or the
  cumulative sum of X-ray point sources in this region.

%
%
%
%

\subsubsection{CXOU~J125904.9$+$345136}\label{59_X4}

CXOU~J125904.9$+$345136 is very weakly detected by {\sl Chandra}.  Its
location is marginally consistent with a bright extended optical
counterpart $\approx$0\farcs4 away, with magnitudes of $m_{\rm
  F814W}=22.1$ and $m_{\rm F547M}=23.9$. Based on the {\sl HST} WFC3
F658N image, this objects lacks any significant rest-frame H$\alpha$
emission at the redshift of Mrk~59, however, which suggests that this
object is likely a background AGN. 
However, an X-ray binary in the tail of Mrk~59 cannot be ruled out.

\subsubsection{CXOU~J125903.9$+$345315}\label{59_X5}

CXOU~J125903.9$+$345315 is only weakly detected at X-ray wavelengths
and is spatially coincident with a $m_{i}\sim25.4$ object which is
marginally detected at the limit of the SDSS imaging. 
Comparison of {\sl Chandra} and {\sl
  XMM-Newton} fluxes shows no variabilty over a time span of $\sim$10 yr,
within the uncertainties. 
This could be a
background object, a foreground star, or a possible X-ray binary in
the trailing tail of Mrk 59.

\subsubsection{CXOU~J125903.9$+$345353}\label{59_X6}

CXOU~J125903.9$+$345353 is detected by both {\sl Chandra} and {\sl
  XMM-Newton}, and is spatially coincident with a $m_{i}\sim19.7$
point-like object that is clearly seen in both SDSS and WISE
imaging. The resulting optical through mid-infrared spectral energy
distribution of this counterpart is nearly constant in flux ($\propto
\nu^{0.1}$) between 4000\AA{} and 12$\mu$m, indicative of either
non-thermal or multiple emission processes. The Palomar H$\alpha$
imaging of \citet{GildePaz2003} shows hints
of rest-frame H$\alpha$ emission at the redshift of Mrk~59, at the
1--2$\sigma$ level in the vicinity of this object, but this could just
be due to a poor continuum subtraction. Finally, the X-ray flux is
seen to vary by a factor of 4--5 between the {\sl XMM-Newton} and {\sl
  Chandra} observations. The above properties suggest that the counterpart
is likely a background AGN, but a spectrum is needed to confirm this.

\subsection{Mrk 71}

\subsubsection{XMMU~J072858.2+691134}\label{71_X1}

XMMU~J072858.2+691134 is the brightest source in Mrk 71. Comparing the
{\it XMM} X-ray contours with the {\it HST} WFC3 data in
Figure~\ref{fig:Mrk71_cutouts}, we see that the X-ray source is aligned
to within $\approx$0\farcs1 with an extended object with $m_{\rm
  F814W}=19.7$. It appears to have a
face-on spiral morphology. However, it does not show any rest-frame
H$\alpha$ emission at the redshift of Mrk~71. This suggests
 that it is likely a background galaxy hosting an AGN.

\begin{figure}
\vspace{0.1in}
\centerline{
\includegraphics[height=9cm, angle=-90]{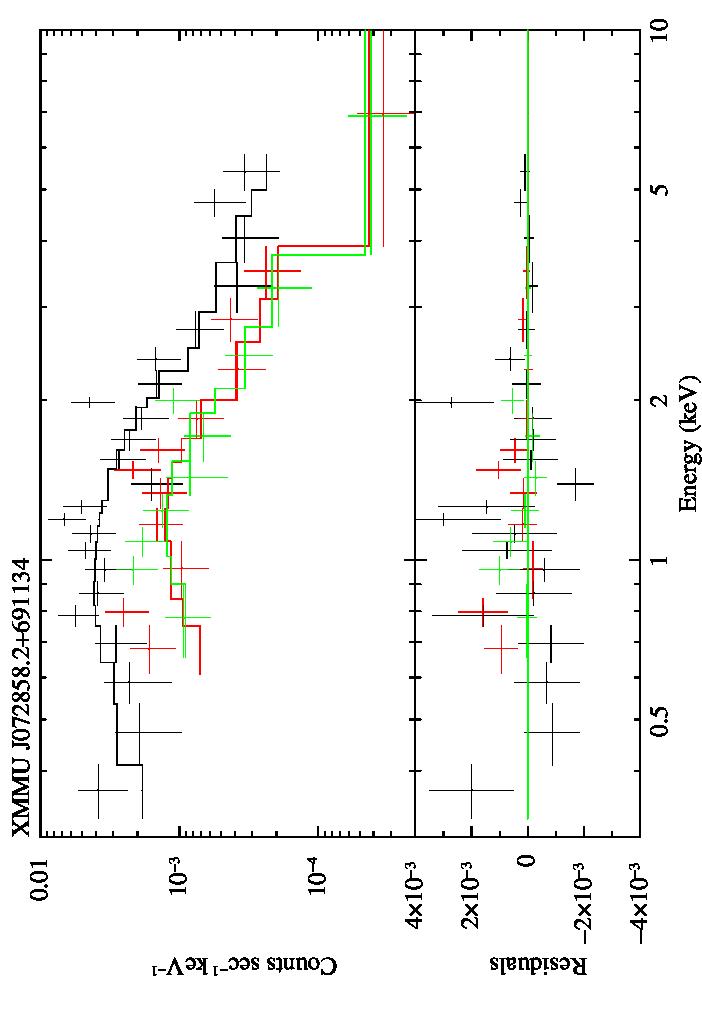}
}
\vspace{0.1cm} 
\caption[]{ X-ray spectra of the brightest X-ray source in Mrk 71, 
XMMU J072858.2+691134.
{\sl XMM-Newton} $pn$, MOS1, and MOS2 spectra are shown in 
black, red and green, 
respectively. The upper panel of each plot shows
  the X-ray spectra and best-fit models (see $\S$\ref{individual} for
  details), while the lower panel shows the residuals of the fits
  for the best-fit model.\label{fig:Mrk71_X1_spectra}}
\vspace{0.2cm} 
\end{figure}

We carried out spectral fitting of the {\sl XMM-Newton} data for this
object. Our choice of the simple power-law 
model [{\tt tbabs*ztbabs*(pow)}] was guided 
by the limited number of counts and the likely optical
identification above.  Again, we fixed the Galactic neutral hydrogen
column density to its fiducial value and allowed the intrinsic column
density and power-law slope $\Gamma$ and normalization to vary. The
spectra were acceptably fitted by this model with parameter values of
$N_{\rm H, intr}=(1.3^{+0.2}_{-0.2})\times10^{21}$~cm$^{-2}$ and
$\Gamma=1.79^{+0.06}_{-0.06}$, with a $\chi^{2}=40.00$ for 42 degrees of
freedom (see Figure~\ref{fig:Mrk71_X1_spectra}). 
The observed 0.5--10.0 keV flux is given in 
  Table \ref{tab:Mrk71}, but we do not calculate a
luminosity because of the unknown redshift of the likely galaxy/AGN
identification.

\subsubsection{XMMU~J072855.4$+$691305}\label{71_X2}

XMMU~J072855.4$+$691305 is the second brightest X-ray source in Mrk
71. Overlaying the {\it XMM} X-ray contours over the {\it HST} WFC3 images 
in Figure~\ref{fig:Mrk71_cutouts}, we see that the X-ray source is
aligned to within $\approx$0\farcs8 with a point-like object with $m_{\rm
  F814W}=17.6$. This object lies just off the {\sl HST} WFPC2
F656N image, but happens to be covered by the CFHT H$\alpha$ image
shown as Fig. 1 in \citet{D00} and by the Kapteyn Telescope
H$\alpha$ survey of \citet{James2004}. In the Kapteyn Telescope image
(rightmost panel of top cutout in Figure~\ref{fig:Mrk71_cutouts}), the
source is spatially coincident within $\approx$1$''$ with a bright, very
compact H {\sc ii} region. This object is also strongly detected by WISE
at 3.6$\mu$m and 4.5$\mu$m. At the distance of Mrk~71, it 
would have an uncorrected absolute magnitude of $M_{\rm F814W}=-10.1$,
placing it among the brightest stars known \citep[certainly much
  brighter than the nearby Luminous Blue Variable NGC 2861 V1, also
  detected in Mrk~71,][]{D00}. Within the uncertainty of the 
{\sl XMM-Newton} position, however, there are many other objects several
magnitudes fainter ($m_{\rm F814W}\ga24.3$), and thus this
identification is very intriguing, but not unique. Another alternative
is that this point-like object is a foreground star, although the
spatial coincidence of such a bright (rare) object with a compact
H {\sc ii} region in Mrk~71 seems improbable. The {\sl XMM-Newton} data
are not sufficient to characterize the spectral nature of the X-ray
emission strongly. With an X-ray luminosity of $6\times10^{36}$ erg
s$^{-1}$, the object would be comparable to Galactic accreting or
colliding-wind HMXBs.

There is another faint potential X-ray point source within the
field-of-view of the cutout (XMMU~J072851.7$+$691255), traced out by
the X-ray contours. It is not formally detected, possibly due to its
marginal blending with XMMU~J072855.4$+$691305 itself. This object
also lies just off the {\sl HST} WFPC2 F656N image, but is in the field of 
view of 
both H$\alpha$ images from \citet{D00} and \citet{James2004}. It is seen 
to be
spatially coincident with a bright, compact H {\sc ii} region, and thus
could be associated with either another faint X-ray binary or extended
thermal plasma in Mrk~71. This potential X-ray source has no
outstanding optical counterpart, but overlaps with
many dozens of faint stars with $m_{\rm F814W}\sim 23$--24.

\subsubsection{XMMU~J072843.4$+$691123}\label{71_X3}

XMMU~J072843.4$+$691123 is detected and resolved by {\sl XMM-Newton},
with an extent of $\approx7.3\pm1.6\arcsec$ as measured by the
3XMM-DR4 catalog \citep{Watson2009}. Given the low number of counts,
  it remains unclear whether it is truly extended or comprised of a
  few point sources.  Comparing the {\it XMM-Newton} position with the
  {\it HST} ACS data in Figure~\ref{fig:Mrk71_cutouts}, we see that
  the extent of the X-ray source aligns well with a dense stellar
  cluster and massive H {\sc ii} complex in the head of Mrk~71.

\subsubsection{XMMU~J072830.4$+$691132}\label{71_X4}

XMMU~J072830.4$+$691132 is very marginally detected by {\sl
    XMM-Newton}. Given the low number of counts, it remains unclear
  whether it is extended or point-like.  Comparing the {\it
    XMM-Newton} position with the {\it HST} ACS data in
  Figure~\ref{fig:Mrk71_cutouts}, we see that the extent of the X-ray
  source again aligns well with a dense stellar cluster and a massive
  H {\sc ii} complex, $\approx$1\farcm1 to the west of the main body of
  Mrk~71.

\begin{table*}
{\tiny
 \begin{minipage}{170mm}
\caption{X-ray Sources Spatially Coincident with Mrk 59}\label{tab:Mrk59} 
  \begin{tabular}{@{}llllrrrrrl@{}}
  \hline
  \hline
{(1)} & 
{(2)} & 
{(3)} & 
{(4)} & 
{(5)} & 
{(6)} & 
{(7)} & 
{(8)} & 
{(9)} & 
{(10)} \\
{Source Name} & 
{RA} & 
{Dec} & 
{Detector} &
{Counts} & 
{HR} & 
{$F_{\rm X}$} & 
{$L_{\rm X}$} & 
{$m_{\rm I}$} & 
{Comments} \\
\hline
\hline
\multicolumn{10}{c}{Mrk 59}\\
\hline
\hline
CXOU~J125900.3$+$345043              & 12 59 00.33 & $+$34 50 42.9 & ACIS-I &   $14.5^{ +5.0}_{ -3.8}$ & $-0.88^{+0.15}_{-0.07}$ &  0.5  &  1.1 & ---  & possibly extended \\
\hline
                                     &             &               & $pn$   &  $541.1^{+24.7}_{-23.7}$ & $-0.71^{+0.02}_{-0.02}$ &       &      &      & \\
XMMU~J125900.7$+$345048 (IXO 72, E1) & 12 59 00.81 & $+$34 50 48.2 & MOS1   &  $198.1^{+15.3}_{-14.3}$ & $-0.66^{+0.04}_{-0.04}$ & 11.4  & 21.5 & 23.0 & $M_{\rm I}=-7.1$; point source\\
                                     &             &               & MOS2   &  $180.9^{+14.7}_{-13.6}$ & $-0.66^{+0.04}_{-0.04}$ &       &      &      & \\ 
\vspace{-0.15cm} & \\
                                     &             &               & $pn$   &  $147.5^{+13.3}_{-12.3}$ & $-0.73^{+0.04}_{-0.04}$ &       &      &      & \\
XMMU~J125900.7$+$345048 (IXO 72, E2) &             &               & MOS1   &  $100.5^{+11.3}_{-10.3}$ & $-0.64^{+0.06}_{-0.06}$ & 10.2  & 19.2 & "    & \\
                                     &             &               & MOS2   &  $ 88.2^{+10.6}_{- 9.6}$ & $-0.72^{+0.06}_{-0.05}$ &       &      &      & \\
\vspace{-0.15cm} & \\
                                     &             &               & $pn$   &  ---                     & ---                     &       &      &      & \\
XMMU~J125900.7$+$345048 (IXO 72, E3) &             &               & MOS1   &  $ 99.3^{+11.6}_{-10.5}$ & $-0.69^{+0.06}_{-0.06}$ &  9.8  & 18.5 & "    & \\
                                     &             &               & MOS2   &  $ 92.3^{+11.2}_{-10.1}$ & $-0.69^{+0.07}_{-0.06}$ &       &      &      & \\
\vspace{-0.15cm} & \\
CXOU~J125900.9$+$345048 (IXO 72, E4) & 12 59 00.83 & $+$34 50 47.7 & ACIS-I &  $225.4^{+16.1}_{-15.0}$ & $-0.50^{+0.04}_{-0.04}$ & 11.0  & 20.7 & "    & \\
\hline
                                     &             &               & $pn$   &  $959.1^{+32.3}_{-31.3}$ & $-0.67^{+0.02}_{-0.02}$ &       &      &      & \\
XMMU~J125901.7$+$345115 (IXO 73, E1) & 12 59 01.83 & $+$34 51 13.9 & MOS1   &  $313.1^{+18.9}_{-17.9}$ & $-0.57^{+0.03}_{-0.03}$ & 23.0  & 38.9 & 21.7 & $M_{\rm I}=-8.5$; point source\\
                                     &             &               & MOS2   &  $314.9^{+18.9}_{-17.9}$ & $-0.57^{+0.03}_{-0.03}$ &       &      &      & \\
\vspace{-0.15cm} & \\
                                     &             &               & $pn$   &  $228.5^{+16.3}_{-15.2}$ & $-0.61^{+0.04}_{-0.04}$ &       &      &      & \\
XMMU~J125901.7$+$345115 (IXO 73, E2) &             &               & MOS1   &  $143.5^{+13.2}_{-12.2}$ & $-0.52^{+0.06}_{-0.05}$ & 18.3  & 31.1 & "    & \\
                                     &             &               & MOS2   &  $145.2^{+13.2}_{-12.2}$ & $-0.62^{+0.05}_{-0.05}$ &       &      &      & \\
\vspace{-0.15cm} & \\
                                     &             &               & $pn$   &  ---                     & ---                     &       &      &      & \\
XMMU~J125901.7$+$345115 (IXO 73, E3) &             &               & MOS1   &  $128.3^{+12.9}_{-11.8}$ & $-0.60^{+0.06}_{-0.05}$ & 14.1  & 23.8 & "    & \\
                                     &             &               & MOS2   &  $127.3^{+12.8}_{-11.7}$ & $-0.62^{+0.06}_{-0.05}$ &       &      &      & \\
\vspace{-0.15cm} & \\
CXOU~J125901.8$+$345115 (IXO 73, E4) & 12 59 01.81 & $+$34 51 14.5 & ACIS-I &  $915.1^{+31.3}_{-30.2}$ & $-0.48^{+0.02}_{-0.02}$ & 52.3  & 88.8 & "    & \\
\hline
CXOU~J125904.8$+$345136              & 12 59 04.82 & $+$34 51 36.3 & ACIS-I &    $6.7^{ +3.8}_{ -2.6}$ & $ 0.71^{+0.16}_{-0.29}$ &  0.2  & ---  & 22.1 & background AGN?\\
%
\hline
                                     &             &               & $pn$   &   $<4.4$                 & ---                     &       &      & \\
XMMU~J125904.0$+$345315 (E1)         & 12 59 04.02 & $+$34 53 14.5 & MOS1   &   $10.3^{ +5.1}_{ -4.0}$ & $<-0.49$                &  0.5  & ---  &  25.4 & background AGN?\\
                                     &             &               & MOS2   &   $<6.3$                 & ---                     &       &      &      & \\
\vspace{-0.15cm} & \\
                                     &             &               & $pn$   &   $ 7.9^{ +4.6}_{ -3.4}$ & $-0.35^{+0.37}_{-0.32}$ &       &      &      & \\
XMMU~J125904.0$+$345315 (E2)         &             &               & MOS1   &   $12.5^{ +5.1}_{ -4.0}$ & $-0.38^{+0.26}_{-0.24}$ &  0.7  & ---     & "    & \\
                                     &             &               & MOS2   &   $<6.5$                 & ---                     &       &      &      & \\
\vspace{-0.15cm} & \\
                                     &             &               & $pn$   &   ---                    & ---                     &       &      &      & \\
XMMU~J125904.0$+$345315 (E3)         &             &               & MOS1   &   $15.7^{ +6.1}_{ -5.0}$ & $<-0.29$                &  1.8  & ---     & "    & \\
                                     &             &               & MOS2   &   $18.8^{ +6.4}_{ -5.3}$ & $-0.44^{+0.22}_{-0.21}$ &       &      &      & \\
\vspace{-0.15cm} & \\
CXOU~J125903.9$+$345315 (E4)             & 12 59 03.90 & $+$34 53 15.0 & ACIS-I &   $ 9.9^{ +4.3}_{ -3.1}$ & $<-0.84$                &  0.4  & ---  & "    & \\
%
\hline
                                     &             &               & $pn$   &   $25.7^{ +7.5}_{ -6.5}$ & $<-0.62$                &       &      &      & \\
XMMU~J125903.9$+$345353 (E1)         & 12 59 03.96 & $+$34 53 53.4 & MOS1   &   $ 7.3^{ +4.7}_{ -3.6}$ & $<-0.33$                &  0.5  & ---  & 19.7 & background AGN?\\
                                     &             &               & MOS2   &   $ 9.5^{ +4.8}_{ -3.7}$ & $-0.29^{+0.33}_{-0.30}$ &       &      &      & \\
\vspace{-0.15cm} & \\
                                     &             &               & $pn$   &   $11.9^{ +5.1}_{ -4.0}$ & $-0.57^{+0.25}_{-0.21}$ &       &      &      & \\
XMMU~J125903.9$+$345353 (E2)         &             &               & MOS1   &   $11.5^{ +5.0}_{ -3.8}$ & $-0.32^{+0.27}_{-0.26}$ &  0.7  & ---  & "    & \\
                                     &             &               & MOS2   &   $ 5.9^{ +4.1}_{ -2.9}$ & $< 0.19$                &       &      &      & \\
\vspace{-0.15cm} & \\
                                     &             &               & $pn$   &   ---                    & ---                     &       &      &      & \\
XMMU~J125903.9$+$345353 (E3)         &             &               & MOS1   &   $ 6.7^{ +5.1}_{ -4.0}$ & $-0.53^{+0.13}_{-0.40}$ &  0.6  & ---  & "    & \\
                                     &             &               & MOS2   &   $ 8.8^{ +5.3}_{ -4.2}$ & $ 0.19^{+0.38}_{-0.42}$ &       &      &      & \\
\vspace{-0.15cm} & \\
CXOU~J125903.9$+$345353 (E4)             & 12 59 03.92 & $+$34 53 53.6 & ACIS-I &   $92.9^{+10.7}_{ -9.6}$ & $-0.38^{+0.07}_{-0.07}$ &  3.3  & ---  & "    & \\
\hline
\end{tabular}
{\sc Notes:} -- 
Column 1: Source name given as CXOU~JHHMMSS.S$+$DDMMSS or XMMU~JHHMMSS.S$+$DDMMSS.
Column 5: Background-subtracted 0.5-8.0~keV counts for {\it Chandra}
ACIS instrument and for the {\it XMM-Newton} $pn$ and MOS
instruments. The errors for the source and background counts were
computed following the method of \citet{Gehrels1986} and were then
combined following the ``numerical method'' described in $\S$1.7.3 of
\citet{Lyons1991}. Sources denoted by '---' were strongly affected by chip gaps.
Column 6: Hardness ratios, defined in a similar manner to those of the
{\it Chandra} \citep{Evans2010} and XMM Serendipitous Source Catalogs
\citep{Watson2009} as the ratio of the difference between the
hard-band (2--8 keV) and soft-band (0.5-2 keV) counts over the sum of
the hard- and soft-band counts.  The quoted hardness ratios have been
corrected for differential vignetting between the hard band and soft
band using the appropriate exposure maps. Errors for this quantity are
calculated following the ``numerical method'' described in Sect. 1.7.3
of \citet{Lyons1991}. Hardness ratios could not be
calculated for sources denoted by '---'. For reference, a $\Gamma=2$ power law absorbed by Galactic
$N_{\rm H}$ would have hardness ratios of $-0.62$, $-0.52$, and
$-0.41$ with the $pn$, MOS, and ACIS-I instruments.
Column 7: Observed, aperture-corrected 0.5--10~keV fluxes in units of
10$^{-14}$~erg~cm$^{-2}$~s$^{-1}$ from the best-fitted models to the
X-ray spectra. For faint sources ($\la$ 100 counts), fluxes were
calculated assuming an average absorbed power-law spectrum with $N_{\rm
  H} = 3.2\times10^{21}$~cm$^{-2}$ and $\Gamma=2.0$.
Column 8: Absorption-corrected 0.5--10~keV luminosities in units of
10$^{38}$~erg~s$^{-1}$ from the best-fit models to the X-ray spectra.
Column 9: Aperture-corrected F814W (``$I$'') Vega magnitudes for
sources imaged by {\it HST}; otherwise SDSS $i$ catalog
magnitudes. See details on individual sources in $\S$\ref{individual}.
\end{minipage}
}
\end{table*}

\begin{table*}
{\tiny
 \begin{minipage}{170mm}
\caption{X-ray Sources Spatially Coincident with Mrk 71}\label{tab:Mrk71} 
  \begin{tabular}{@{}llllrrrrrl@{}}
  \hline
  \hline
{(1)} & 
{(2)} & 
{(3)} & 
{(4)} & 
{(5)} & 
{(6)} & 
{(7)} & 
{(8)} & 
{(9)} & 
{(10)} \\
{Source Name} & 
{RA} & 
{Dec} & 
{Detector} &
{Counts} & 
{HR} & 
{$F_{\rm X}$} & 
{$L_{\rm X}$} & 
{$m_{\rm I}$} & 
{Comments} \\
\hline
                                     &             &               & $pn$ &  $ 16.0^{ +6.8}_{ -5.7}$ & $<0.22$                 &       &      &      & \\
XMMU~J072830.4$+$691132              & 07 28 30.36 & $+$69 11 31.9 & MOS1 &  $  4.1^{ +4.8}_{ -3.7}$ & ---                     &  0.14  & 0.03 &      & extended? \\
                                     &             &               & MOS2 &  $  7.8^{ +6.5}_{ -5.4}$ & $-0.22^{+0.73}_{-0.69}$ &       &      &      & \\
\hline
                                     &             &               & $pn$ &  $193.4^{+16.2}_{-15.1}$ & $-0.30^{+0.06}_{-0.05}$ &       &      &      & \\
XMMU~J072858.2$+$691134              & 07 28 58.18 & $+$69 11 33.6 & MOS1 &  $108.5^{+12.4}_{-11.3}$ & $-0.39^{+0.07}_{-0.07}$ &  3.4  & ---  & 19.7 & background galaxy/AGN \\
                                     &             &               & MOS2 &  $ 94.8^{+11.8}_{-10.8}$ & $-0.35^{+0.08}_{-0.08}$ &       &      &      & \\
\hline
                                     &             &               & $pn$ &  $ 35.4^{ +9.5}_{ -8.4}$ & $-0.81^{+0.15}_{-0.15}$ &       &      &      & \\
XMMU~J072843.4$+$691123              & 07 28 43.38 & $+$69 11 23.1 & MOS1 &  $  6.1^{ +5.1}_{ -4.0}$ & $< 0.02$                &  0.3  & 0.08 &      & extended \\
                                     &             &               & MOS2 &  $ 12.1^{ +5.9}_{ -4.8}$ & $-0.60^{+0.29}_{-0.27}$ &       &      &      & \\
\hline
                                     &             &               & $pn$ &  $ 43.4^{ +9.9}_{ -8.9}$ & $-0.84^{+0.12}_{-0.12}$ &       &      &      & \\
XMMU~J072855.4$+$691305              & 07 28 55.45 & $+$69 13 05.0 & MOS1 &  $ 23.5^{ +7.7}_{ -6.6}$ & $<-0.69$                &  0.5  & 0.06 & 17.6 & $M_{\rm I}=-10.1$; point source\\
                                     &             &               & MOS2 &  $ 16.8^{ +7.2}_{ -6.1}$ & $<-0.77$                &       &      &      & \\
\hline
\end{tabular}

{\sc Notes:} -- 
Same as Table 1.
\end{minipage}
}
\end{table*}

\section {Discussion} 

\subsection{Mrk 59}

As seen in Figure~\ref{fig:overlay1}, Mrk 59 exhibits two very bright
point sources, with 0.5--10 keV luminosities of $\approx$2$\times$10$^{39}$
erg s$^{-1}$ and $\approx$(2--9)$\times$10$^{39}$ erg s$^{-1}$ for IXO 72
and 73, respectively.
In addition, there is faint emission around the H {\sc ii} complex
constituting the cometary ``head'' which amounts to a few percent of the
total X-ray emission. The two bright X-ray objects are spatially
coincident with compact H {\sc ii} regions, suggesting they are
associated with HMXBs or supernovae. Their
X-ray spectral and variability properties favor the HMXB hypothesis in
both cases. As such, their proximity to compact H {\sc ii} regions also
implies that both they and their counterparts are quite young and
potentially quite massive. Notably, IXO 73 is spatially coincident
with an uncorrected $M_{\rm F814W}=-7.1$ star. 
Because the 2 X-ray sources appear to be associated with 
single objects, their luminosities 
place them in the range of the so-called ultraluminous X-ray sources
\citep[ULXs; e.g. ][]{M00}.

The derived best-fit H {\sc i} column density for both IXO 72 and IXO
73 are somewhat lower than the adopted peak H {\sc i} column density
read off the 21 cm maps of \citet{THL04}. This may imply that both
objects reside either on the near side of the galaxy or within a large
low-density cavity. The upper limits for both sources are still within
30--50\% of the peak value, however, and could be consistent with
these maps: the radio beam size being $\sim$30\arcsec, a X-ray source
which is slightly spatially offset from the position of the peak H
{\sc i} column density would be associated with a lower column
density.

ULXs are rare. Most galaxies, including our own Milky Way, have none, and 
galaxies that do host a ULX usually have only one. Mrk 59 possesses two. 
In a {\sl Chandra} study of three of the most 
metal deficient BCDs known in the local universe, SBS 0335--052W, I Zw 18 and 
SBS 0335--052E, with 12 + log O/H = 7.12 (2.6\% $Z_\odot$), 7.17 
(2.9\% $Z_\odot$)and 7.31 (4.0\% $Z_\odot$), respectively, \citet{T04} found 
also that more 
than 90\% of the X-ray emission of these 3 BCDs arise from point-like sources,
and that the 0.5--10 keV luminosities of these point sources are in the
ULX domain, with X-ray luminosities in the 
(1.3--8.5)$\times$10$^{39}$ erg s$^{-1}$ range.  Both SBS 0335--052E and 
I Zw 18 contains one ULX, while SBS 0335 --052W contains two.

 
\citet{T04} suggested that 
the high X-ray luminosities
of HMXBs in BCDs could be the result of their lower metallicities. 
The conclusion that ULXs occur preferentially in lower-metallicity
systems has been recently strengthened by the work of \citet{P13} who
observed with {\sl Chandra} 22 more low-metallicity (12 + log O/H $<$
7.65 or $<$ $Z_\odot$/11) objects, the majority of them being
BCDs. They detected 3 more objects, with X-ray luminosities again in
the range of ULXs. Adding the X-ray detections
of SBS 0335--052W, I Zw 18 and SBS 0335--052E by \citet{T04},  \citet{P13} 
constructed a low-metallicity galaxy sample which
included a total of 6 X-ray detections. They then compare the low-metallicity 
sample to a control galaxy sample with higher metallicities, and  
conclude
that ULXs occur preferentially in the metal-poor sample with a formal
statistical significamce of 2.3 $\sigma$. The detection of the 2 
ULXs in Mrk 59 brings to 7 the number of low-metallicity BCDs known to contain such objects. 

\subsection{Mrk 71}

Figure~\ref{fig:overlay2} shows the four detected X-ray sources
associated with Mrk~71. The strongest two objects are point-sources 
and have relatively
clear optical counterparts. We identify the brightest with a
background galaxy/AGN, while the second brightest coincides with an
exceptionally bright point-like object and compact H {\sc ii} region,
suggesting it may be an extremely luminous and massive star. The
properties from the limited X-ray data appear consistent with these
identifications. The two remaining X-ray sources are 
faint and potentially extended, and 
appear to be associated with high surface brightness H~{\sc ii}
complexes at the end of the stellar body, regions I and III in the
notation of \citet{D00}. {\sl HST} imaging of region I by \citet{D00}
and \citet{TI05} shows that it contains two young compact clusters, A
and B. A Luminous Blue Variable (LBV) star was also discovered by
\citet{D00} in region I. 
The spatial resolution of the X-ray map does not allow to
say whether the X-ray source in region I, XMMU~J072843.4$+$691123, 
is associated with the overall
H {\sc ii} complex or individual objects, such as the LBV star. 

\subsection{ULXs and low-metallicity environments}

How may one understand the observed statistical 
increase of ULXs in low-metallicity BCDs? 
Three main scenarios have been proposed to account for the very high luminosity 
of ULXs (see e.g. \citet{Z09} and references therein). A first scenario 
associates  
ULXs with stellar mass black holes (BH), 
with masses less than about 20 $M_\odot$. Their high X-ray luminosities 
come from either 
anisotropic emission (or beaming) or super-Eddington accretion   
via a massive accretion disc, the structure of which has been 
somehow modified, or 
both. In the second scenario, the compact object is considerably more massive,
being an intermediate-mass black hole (IMBH) with mass in excess of 100 $M_\odot$ and up to thousands $M_\odot$.    
The accretion rate would not be in the super-Eddington, but 
in the usual sub-Eddington regime. 
The third scenario is intermediate 
between the first two: it also invokes a stellar mass 
BH, but in a slightly higher mass range, between $\sim$ 30 and 
90 $M_\odot$. To account for the high X-ray luminosity, 
the accretion needs to be super-Eddington (by a factor of a few) and  
modest beaming (a beaming factor of $\sim$ 0.5) is required.
While in the first scenario with beaming and super-Eddington accretion, 
a direct link between low-metallicity and ULXs is not clearly evident,
we can see more of such a link in the second and third scenarios. Metal-poor environments do favour the formation  
of massive stars, and hence of massive black holes \citep[e.g.][]{B04}. 
 Although the second and third scenarios have their pros and cons , \citet{Z09} have presented plausible arguments in favor of  
the third scenario.
At the end of their lives, 
subsolar metallicity stars with masses above $\sim$ 30--40 $M_\odot$ retain
their massive envelopes at the time of explosion because these are not 
removed efficiently through line-driven winds as in solar metallicity stars.
So most of the star collapses to a BH with a mass comparable to that of the 
pre-supernova star \citep{F99}. Their masses would not significantly exceed 
$\sim$80--90 $M_\odot$ because a more massive star would undergo pulsational 
pair-instability in its core and eject most of its envelope. This scenario has 
the advantage over the IMBH scenario in that it does not require a new exotic 
mechanism to produce very massive BHs in starbursts, but simply appeals to 
ordinary stellar evolution coupled with metallicity effects. It is also more 
attractive than the stellar mass 
BH with $M$$\leq$20 $M_\odot$ scenario: with less extreme
beaming factors and violations of the Eddington limit, it does not require 
extreme accretion scenarios.     

  
However, while metallicity plays a important role in determining the X-ray
luminosities $L_X$ of HMXBs, it cannot be the only factor as  
ULXs in the same metallicity environment have different X-ray luminosities.  
Thus, the
E and W components of the SBS 0335--052 system have about the same
metallicities, but $L_X$ of SBS 0335--052W is 2.4 times larger than
that of SBS0335--052E. The X-ray luminosities of the two sources in
Mrk 59 also differ by a factor of $\sim$2.  The age of the burst is likely to
be also a factor. As the timescale for HMXB formation is 3--10 Myr and
the maximum number of HMXBs is expected to occur 20--50 Myr after the
starburst, then we may expect, in a statistical sense, that the HMXB X-ray luminosity depends 
on the age of the starburst.  If we use
the equivalent width of the H$\beta$ emission line EW(H$\beta$) as an
age indicator: the W component has EW(H$\beta$) = 80 \AA,
corresponding to a starburst age of $\sim$4 Myr \citep[using the
  instantaneous star formation model with $Z$ = 0.001, a Salpeter
  slope for the stellar initial mass function and $M_{up}$ = 100
  $M_\odot$ of ][]{L99}, while the E component has EW(H$\beta$) =
190~\AA, corresponding to a younger starburst age of $\sim$3 Myr, and
hence to a lower HMXB X-ray luminosity. 
As for the objects studied here, the X-ray sources in Mrk 71 are considerably 
weaker than those in Mrk 59, although the two BCDs have about the same 
metallicity. 
The
X-ray luminosities of the point-source and  
two extended objects in Mrk 71 are at least 1--2
orders of magnitude lower than that of their counterparts  
in Mrk~59. 
 The high surface
brightness H~{\sc ii} region in Mrk~71 has EW(H$\beta$) = 357\AA\ as
compared to EW(H$\beta$) = 157\AA\ for Mrk~59 \citep{ITL97}.  Using
the instantaneous model of \citet{L99} with the same parameters as
described before, these EW(H$\beta$) correspond to starburst ages of
2.5 Myr and 4 Myr, respectively. Thus the weakness of the
X-ray sources  in Mrk 71 to be due to the fact that
HMXBs have not had time to form, the onset of HMXB formation starting
at $\sim$ 3 Myr.

Admittedly, these are only two suggestive examples. A more extensive study with a larger statistical sample will be needed to demonstrate or eliminate the hypothesis that the HMXB X-ray luminosities are correlated with the age of the 
starburst.

\section{Conclusions}

We have investigated the X-ray emission of the two cometary Blue Compact
Dwarf (BCD) galaxies Mrk 59 and Mrk 71, based on {\sl XMM} and {\sl Chandra}
observations. Our main findings are the following:

1. Mrk 59 contains two very bright X-ray point sources, IXO 72 and IXO 73,
with 0.5 -- 10 keV luminosities of (1.8--2.1)$\times$10$^{39}$ and
(2.4-8.9)$\times$10$^{39}$ erg s$^{-1}$, respectively. The cometary
``head'' H {\sc ii} complex is also faintly detected, its diffuse 
emission constituting a
few percent of the total X-ray emission from the galaxy. Both IXO 72 and 
IXO 73 possess optical counterparts, IXO 72's counterpart being
potentially identified as an individual luminous, massive star while
IXO 73's counterpart is a bright stellar object  
located in an slightly resolved compact H {\sc ii}
region. 
The above identifications suggest that both IXO 72 and IXO 73 are
single objects, thus qualifying them as legitimate ultraluminous X-ray (ULX)
sources. The  0.5 -- 10 keV X-ray flux of IXO 72 has 
remained approximately constant over 
the past 10 yr, while that of IXO 73 has varied by a factor of
$\approx$4 over the same period. 
The X-ray spectra of both sources are typical of ULXs.
Such high X-ray luminosities may be related to the low metallicity of Mrk 59
(0.2 solar). 

2. Mrk 71 contains four faint X-ray sources. The brightest one
is spatially coincident with a background spiral galaxy. The second
brightest one is coincident with a very compact H {\sc ii} region and a
bright star; if associated with Mrk~71, this star is extremely
luminous and among the brightest stars known. The other two faint
X-ray sources are associated with large H {\sc ii} complexes.  
All three sources are 1--2 orders of
magnitude fainter the Mrk 59 X-ray sources. As 
Mrk 71 has the same metallicity as as Mrk 59, metallicity 
cannot be the only factor in determining X-ray luminosities. The age of the 
starburst may play a role. 

\section{Acknowledgements}

T.X.T. acknowledges the support of NASA grant NAG5-12937.  F.E.B.
acknowledges support for this project from Basal-CATA (PFB-06/2007),
CONICYT-Chile (under grants FONDECYT 1141218 and Anillo ACT1101),
Project IC120009 "Millennium Institute of Astrophysics (MAS)", funded
by the Iniciativa Cient\'{\i}fica Milenio del Ministerio de Econom\'{\i}a,
Fomento y Turismo de Chile, and Chandra Postdoctoral Fellowship grant
number PF4-50032 awarded by the Chandra X-ray Center, which is
operated by the Smithsonian Astrophysical Observatory for NASA under
contract NAS8-03060.

This research has made use of 
data obtained from the High Energy Astrophysics Science Archive
Research Center (HEASARC), provided by NASA's Goddard Space Flight
Center and NASA/ESA Hubble Space Telescope, obtained from the data
archive at the Space Telescope Science Institute. STScI is operated by
the Association of Universities for Research in Astronomy, Inc. under
NASA contract NAS 5-26555. This research also made use of software
provided by the Chandra X-ray Center (CXC) in the application package
CIAO and SAOImage DS9, developed by Smithsonian Astrophysical
Observatory. Funding for the SDSS and SDSS-II was provided by the
Alfred P. Sloan Foundation, the Participating Institutions, the
National Science Foundation, the U.S. Department of Energy, the
National Aeronautics and Space Administration, the Japanese
Monbukagakusho, the Max Planck Society, and the Higher Education
Funding Council for England. The SDSS was managed by the Astrophysical
Research Consortium for the Participating Institutions. The Digitized
Sky Surveys were produced at the Space Telescope Science Institute
under U.S. Government grant NAG W-2166. The images of these surveys
are based on photographic data obtained using the Oschin Schmidt
Telescope on Palomar Mountain and the UK Schmidt Telescope. The plates
were processed into the present compressed digital form with the
permission of these institutions.

\bsp

\label{lastpage}

\end{document}